%% file: main.tex
\documentclass[tightenlines,superscriptaddress,10pt,twocolumn,aps,prd]{revtex4-1}

\usepackage{geometry}
\usepackage{graphicx,wrapfig}
\usepackage[export]{adjustbox}
\usepackage{epsfig}
\usepackage{float}
\usepackage{flafter}
\usepackage{subfigure}
\usepackage{sidecap}
\usepackage{notes2bib}
\usepackage{gensymb}
\usepackage{url}
\usepackage{amsmath}
\usepackage{latexsym}
\usepackage{amssymb}
\usepackage{xcolor}
\usepackage{color}
\usepackage{slashed}
\usepackage{siunitx}
\usepackage{notoccite}
\usepackage[utf8]{inputenc}
\usepackage[english]{babel}
\usepackage{csquotes}
\usepackage{dcolumn}
\usepackage{bm}
\usepackage{color}
\usepackage{enumitem}
\usepackage{pifont}
\usepackage{textcomp}
\usepackage{tikz}
\usepackage{colortbl}
\usepackage[version=4]{mhchem}
\usepackage{xspace}
\usepackage{placeins}
\usepackage{tabularx}
\usepackage{hyperref}
\usepackage{amssymb}
\usepackage{hyperref}
\usepackage{lineno}

\usepackage{pgfgantt}
\usepackage[ampersand]{easylist}

\begin{document}

\interfootnotelinepenalty=10000

\twocolumngrid

\title{Accessing new physics with an undoped, cryogenic CsI CEvNS detector for COHERENT at the SNS}

\widowpenalty10000
\clubpenalty10000
\renewcommand\floatpagefraction{1}
\renewcommand\topfraction{1}
\renewcommand\bottomfraction{1}
\renewcommand\textfraction{0}

\setlength{\belowcaptionskip}{-10pt} 

\renewcommand{\thesection}{\arabic{section}}
\renewcommand{\thesubsection}{\thesection.\arabic{subsection}}
\renewcommand{\thesubsubsection}{\thesubsection.\arabic{subsubsection}}

\makeatletter
\renewcommand{\p@subsection}{}
\renewcommand{\p@subsubsection}{}
\makeatother

\input{authors-CryoCsI.tex}

\begin{abstract}
We consider the potential for a 10-kg undoped cryogenic CsI detector operating at the Spallation Neutron Source to measure coherent elastic neutrino-nucleus scattering and its sensitivity to discover new physics beyond the standard model.  Through a combination of increased event rate, lower threshold, and good timing resolution, such a detector would significantly improve on past measurements.  We considered tests of several beyond-the-standard-model scenarios such as neutrino non-standard interactions and accelerator-produced dark matter.  This detector's performance was also studied for relevant questions in nuclear physics and neutrino astronomy, namely the weak charge distribution of CsI nuclei and detection of neutrinos from a core-collapse supernova.  
\end{abstract}

\maketitle

\section{Introduction}

A scintillating CsI[Na] crystal was used for the first detection of coherent elastic neutrino-nucleus scattering (CEvNS)~\cite{PhysRevD.9.1389,COHERENT:2017ipa} at the Spallation Neutron Source (SNS) at Oak Ridge National Laboratory.  This detector achieved a light yield of 13.35 photoelectrons (PE) per keV of electron equivalent energy (keV$_\mathrm{ee}$), which set a threshold nuclear recoil energy of $\approx 8$~keV$_\mathrm{nr}$, allowing detection of the low-energy nuclear recoils produced in CEvNS interactions.  This result has improved precision for measuring the standard model's predicted neutrino couplings and searching for physics beyond the standard model (BSM)~\cite{Papoulias:2017qdn,Coloma:2017ncl,Proceedings:2019mnq,PhysRevD.104.015019,PhysRevD.105.033001,Dutta:2020vop,Miranda:2020tif,Banerjee:2021laz}.  Though successful, first-light CEvNS measurements suffered from limited sample statistics and a relatively high detection threshold.  The next generation of CEvNS detectors must address both of these concerns to fully realize the power of precision CEvNS scattering experiments to discover new physics. 

The use of undoped, inorganic scintillators such as CsI and NaI operated at cryogenic temperatures~\cite{Mikhailik:2014wfa,CLARK20186,Ding:2020uxu,Lee_2022} has been studied recently as a potential improvement.  At 77~K, the light output of such crystals more than doubles compared to doped crystals at room temperatures.  Detectors using this technology have been proposed to search for dark matter~\cite{ANGLOHER201670,PhysRevD.96.016026}, in particular testing the DAMA~\cite{NADEAU201562} result, and precisely measure CEvNS~\cite{Baxter:2019mcx,Su:2023klh,Akimov:2022oyb}. Through a combination of increased light yield and reduction of afterglow, long-lived scintillation activity following a MeV-scale energy deposit that becomes a background for CEvNS analysis, a cryogenic CsI CEvNS detector could achieve a threshold lower than the original CsI threshold by an order of magnitude, making a new and currently untested kinematic region experimentally accessible.  A lower threshold also increases the fraction of CEvNS events that would be selected by data analysis for a detector at a stopped-pion neutrino source.  At a nuclear reactor, such improvement would be necessary for CEvNS detection~\cite{Wang:2022ekc}.  A measurement of CEvNS on multiple targets, light and heavy, is desirable to fully test the standard-model prediction.  Next-generation detectors on argon and germanium, both relatively light, have been either recently deployed at the SNS (Ge) or started construction (Ar).  High-precision data with a heavy nuclear target like CsI would supplement these.

In this work, we describe a broad view of the physics potential of a 10-kg cryogenic CsI detector as part of the COHERENT program at the SNS~\cite{Akimov:2022oyb}. Specifically, we consider three BSM scenarios: tests of neutrino-quark non-standard interactions (NSI), potential to discover hidden-sector dark-matter particles, and searches for a sterile neutrino through BSM neutrino oscillations.  We also study two areas where such data will improve understanding of nuclear and astrophysics: measuring the weak charge distribution of the nucleus and observing neutrinos from a core-collapse supernova.  We refer to this proposed detector as COH-CryoCsI-1 throughout.  Though we considered several of the potential new physics signatures a cryogenic CsI scintillator at the SNS may explore, the list is not exhaustive.  For example, this technology can also test neutrino electromagnetic properties~\cite{Cadeddu:2018dux,AtzoriCorona:2022qrf}, a BSM neutrino magnetic moment~\cite{Giunti:2014ixa,Giunti:2015gga,Lindner:2017uvt,Papoulias:2017qdn,Miranda:2020tif,PhysRevD.107.053001,AtzoriCorona:2022qrf}, and leptoquark models~\cite{DeRomeri:2023cjt}.  

\section{COHERENT program at the SNS}

The SNS is currently the most intense terrestrial source of neutrinos in the 10's of MeV energy range.  During SNS operations, protons are accelerated to a kinetic energy of $T_p=1.01$~GeV and stacked in an accumulator ring.  The protons are then extracted at a rate of 60~Hz and directed to a mercury target.  Each extraction results in a $\approx 350$~ns FWHM pulse.  The proton power upgrade~\cite{Howell_2017} is currently increasing the proton energy to $T_p=1.3$~GeV and beam power to 2.0~MW at the target.  Recently, steady operations at a record 1.7~MW power were achieved.  Work is expected to be complete by mid 2024 with full power. 

The SNS produces $\pi^\pm$ mesons naturally as the proton beam is dumped on the mercury target.  The $\pi^-$ capture in nuclei but the $\pi^+$ will stop in the target and decay freely making a $\pi^+$ decay-at-rest ($\pi$DAR) neutrino flux, $\pi^+\rightarrow\mu^++\nu_\mu$, $\tau=26$~ns.  Subsequently, the $\mu^+$ will then stop in the target and decay, $\mu^+\rightarrow e^++\nu_e+\bar{\nu}_\mu$, $\tau=2.2\mu$s.  Conveniently, the time-scale of the beam is between these two lifetimes so that the flux separates into two components: a prompt, monoenergetic (29.8~MeV) flux of $\nu_\mu$, and a delayed flux of $\nu_e$ and $\bar{\nu}_\mu$ whose energy distributions are very well understood from $\mu^+$ decay kinematics.  At the SNS energy, the flux is a pure $(>99\%)$ $\pi$DAR source with a very small contribution from decay-in-flight mesons~\cite{PhysRevD.106.032003}.  Among artificial neutrino sources, a $\pi$DAR flux produces a large flux of neutrinos in the 10 to 50~MeV energy range.  This energy regime is very useful for astroparticle neutrino measurements~\cite{BAHCALL200047,Hirata:1988ad}; $^8$B solar neutrinos have a 15-MeV endpoint energy while supernova neutrinos have a mean energy of 10-20~MeV.

COHERENT operates a suite of neutrino and background detectors in ``Neutrino Alley", a basement utility hallway where beam-related neutron backgrounds are measured to be small enough to facilitate low-rate neutrino measurements.  Taking full advantage of our low-background environment and the unique energy and intensity of the SNS, we have adopted a multi-target approach, measuring both CEvNS and inelastic neutrino-interaction cross sections.  

Beyond the first CsI[Na] detector, three COHERENT CEvNS detectors are currently operating: COH-Ar-10, a 24-kg argon scintillation calorimeter which has seen 3.4~$\sigma$ evidence for CEvNS~\cite{COHERENT:2020iec};  COH-NaI-3500, currently 1500 kg of NaI scintillating crystals which will measure CEvNS on $^{23}$Na and $\nu_e$~CC interactions on $^{127}$I; and COH-Ge-1, an 18-kg germanium p-type point-contact (PPC) detector array~\cite{PhysRevLett.106.131301,LEGEND:2017cdu,CDEX:2016tve,TEXONO:2014eky,CONUS:2020skt}.  Also taking data are COH-NaI-185~\cite{COHERENT:2023ffx}, which has measured inelastic neutrino interactions on NaI; COH-Th-1, which will measure the first neutrino cross section on thorium~\cite{Jordan_IV_2004,Qian:2002mb,Kolbe:2003jf}; COH-D2O-1, which will calibrate the neutrino flux by studying $\nu_e-d$ interactions~\cite{Akimov_2021}; and MARS, a neutron background monitor~\cite{COHERENT:2021qbu}.

\section{Cryo CsI detector performance}
\label{sect:Detector}

\begin{figure}
  \includegraphics[width=0.48\textwidth]{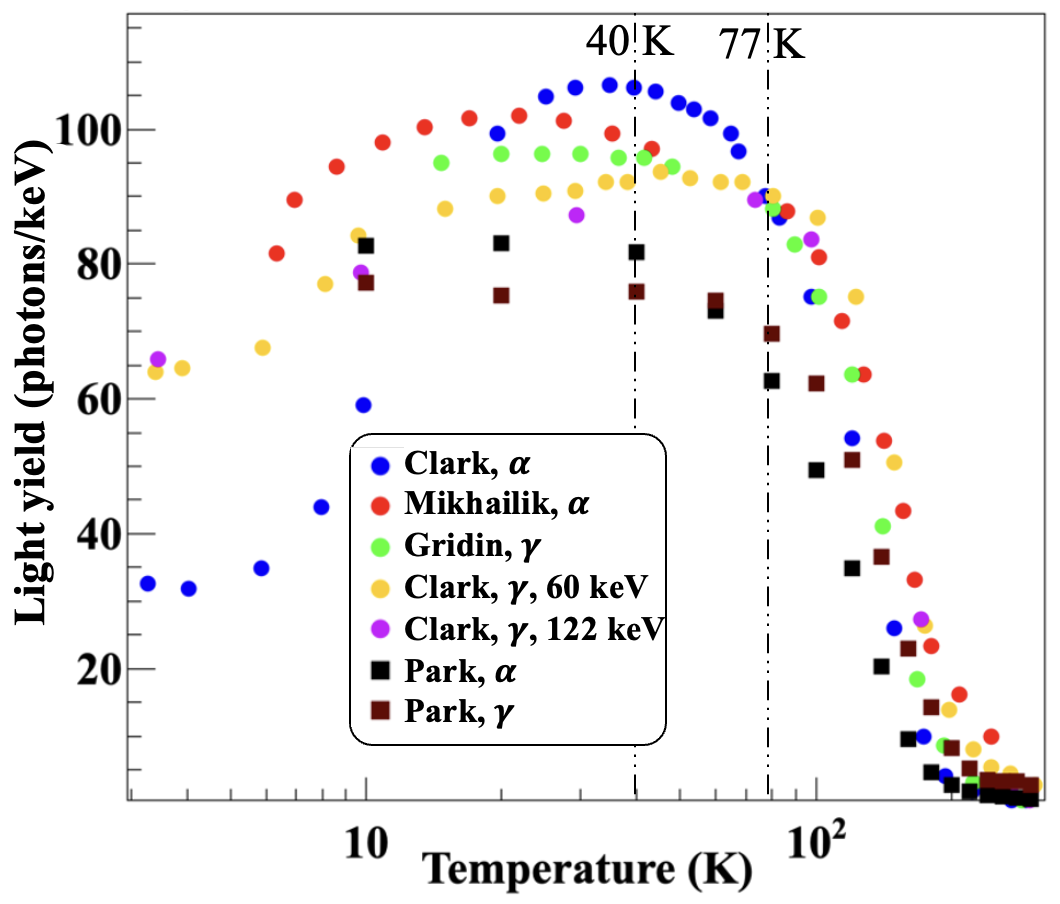}
  \caption{The light yield as a function of operating temperature for undoped CsI scintillators.  For undoped CsI, $\sim$~40~K is an ideal temperature.  Figure reproduced from~\cite{Park:2021jwv}.  Data from~\cite{PhysRevB.51.2167,Sailer_2012} based on alpha and gamma irradiation.  Quenching is accounted for with the alpha measurements.}
  \label{fig:LYAfterglow}
\end{figure}

The COHERENT CsI detector that first observed CEvNS achieved a light yield of 13.35~PE/keV$_\mathrm{ee}$, but it was only able to achieve a threshold of $\approx$~700~eV$_\mathrm{ee}$ due to a 9~PE coincidence cut to remove both Cherenkov light in the photomultiplier tube (PMT) and the prominent afterglow observed in doped CsI[Na] crystals~\cite{Collar:2014lya} at room temperature.  There are three strategies to improve threshold relative to the original CsI detector: switch from PMT to silicon photomultiplier (SiPM) light detectors, reduce the afterglow scintillation rate, or increase the light yield. By switching to a SiPM readout for COH-CryoCsI-1, all three of these will be simultaneously met for undoped CsI crystals operating near 40~K where light yield is optimized, as shown in Fig.~\ref{fig:LYAfterglow}.  

COHERENT collaborators performed several tests of the light yield of undoped, cryogenic inorganic scintillators~\cite{Chernyak:2020lhu,Ding:2020uxu,Ding:2022jjm} at the University of South Dakota.  From these, we have preliminarily estimated the expected COH-CryoCsI-1 detector performance.  SiPM light detectors yield the most favorable light collection, primarily due to their large 40-50$\%$ quantum efficiency~\cite{Jackson2014HighvolumeSP}.  In these tests, the light yield of a small cube of undoped CsI, outfitted with a SiPM array on two faces and operated at 77~K, was measured as $43.0\pm1.1$~PE/keV$_\mathrm{ee}$~\cite{Ding:2022jjm}.  Yet-unpublished results have achieved $>50$~PE/keV$_\mathrm{ee}$ using a wavelength-shifting paint in the setup.  The dark-count rate per active SiPM area was also measured at this temperature and used in a  Monte Carlo which showed a 10~$\mu$Hz trigger rate after requiring a coincidence, $\Delta t < 10$~ns, of two PE pulses observed in different SiPM arrays in the 10~kg detector.  When coupled with the small duty factor of the SNS beam, $\approx30\times10^{-5}$, this gives a negligible $<1$ selected dark count per year from the SiPM assembly~\cite{Ding:2022jjm}.  

These results, extrapolated from 77~K to 40~K operations, would suggest a noticeably higher light yield.  For this work, we assume a light yield of 50~PE/keV$_\mathrm{ee}$, the highest measured at a test stand.  Meanwhile, the dark count rate in the SiPM arrays will be lower at a lower crystal temperature, as will the afterglow rate in CsI~\cite{Derenzo:2018plr}.  Combined, we expect a $\approx$~4~PE threshold for the final detector after requiring a 2~PE coincidence for selection and considering the scintillation timing of the crystal, equivalent to 80~eV$_\mathrm{ee}$.  

The COH-CryoCsI-1 shielding will be based on COHERENT's first CsI detector~\cite{COHERENT:2017ipa} which used both lead and high-density polyethylene to eliminate gamma and moderate beam-correlated neutron backgrounds.  This shielding was very effective for the initial detector. Background levels below 18.7~keV$_\mathrm{ee}$ were monitored in a 12-$\mu$s interval prior to the arrival of the beam signal both during beam-on and beam-off data collection.  The difference in count rates during beam-on and beam-off operations was $(-0.9\pm1.4)\%$, consistent with no excess due to the increased radioactivity from the SNS.  As the COH-CryoCsI-1 detector will be operated at 40~K, a cryostat is necessary and may introduce radiological backgrounds.  The innermost stage of the cryostat will be constructed of low-activity copper to avoid this.  

\begin{figure}
  \includegraphics[width=0.48\textwidth]{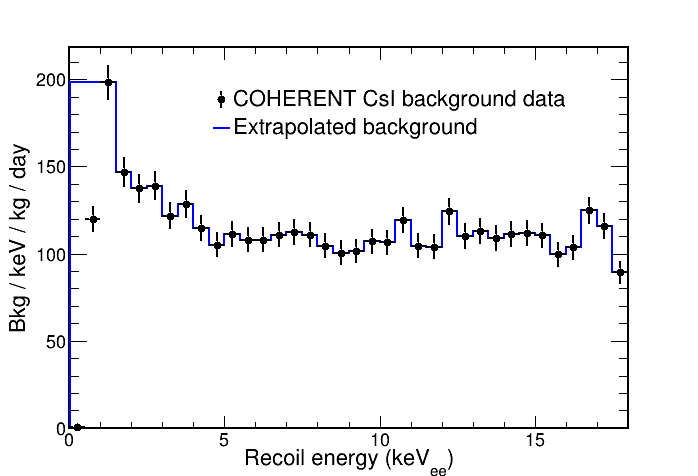}
  \caption{The steady-state background model assumed in COH-CryoCsI-1, based on initial data from COHERENT's first CsI detector.  The first two data points are suppressed due to the original detector's high threshold. 
 These points are increased to 196 counts/keV/kg/day, the value in the third bin.}
  \label{fig:SSBkg}
\end{figure}

In CsI crystals, the principal intrinsic backgrounds are $\beta^-$ decays of $^{87}$Rb, a primordial radionuclide that chemically contaminates the crystals, and $^{137}$Cs.  Attention must be given to the concentration of these impurities in the procured crystal.  The first COHERENT CsI detector had measured concentrations of 92~ppb of $^{87}$Rb and $28\pm3$~mBq/kg of $^{137}$Cs~\cite{Collar:2014lya}.  Preliminary measurements of $^{87}$Rb concentration from the SICCAS crystal vendor show reduced $^{87}$Rb concentrations, $1-4$~ppb.  As such, the backgrounds in COH-CryoCsI-1 may be significantly reduced compared to first CEvNS measurements.  We assume the background rate will be the same as the observed background in the original CsI detector, shown in Fig.~\ref{fig:SSBkg}.  Further, we extrapolate the background rate as constant below 1.5~keV$_\mathrm{ee}$ although background dropped below threshold in CsI data.  We assume that BRN events are the only standard-model backgrounds.  Inelastic neutrino interactions, described in~\cite{Bednyakov2021ConceptOC}, have no overlap with the region of interest for CEvNS, either depositing no energy or a high-energy gamma.  Also noteworthy, the beam power is planned to be be nearly doubled during COH-CryoCsI-1 running (2.0~MW) compared to the average (1.14~MW) delivered by the SNS to the first CsI detector which also improves signal-to-background.  Sensitivity estimates assume the planned 2.0~MW running.

There was also a small number of beam-correlated neutron events recorded within the CsI detector.  The neutron flux is known to vary over the length of Neutrino Alley; we will place COH-CryoCsI-1 in the vicinity of the original CsI detector location.  We thus assume the neutron flux observed in that detector will also be incident on COH-CryoCsI-1.  Correcting for beam power, detector mass, and detector threshold, this contributes $\approx21$ neutron events per year.

As a last but vitally important component required for calculating the CEvNS signal prediction, we must assume a nuclear quenching.  Generally, only a fraction of the energy deposited by a recoiling nucleus produces scintillation light -- much is lost due to ionization and heat.  This quenching factor depends strongly on material, doping, temperature, and other parameters.  COHERENT collaborators have taken quenching factor data with undoped CsI at 77~K at Triangle Universities Nuclear Laboratory (TUNL).  Though data analysis is underway, preliminary estimates point to a roughly energy-independent quenching factor of $\approx15\%$.  We further assume a 10$\%$ relative uncertainty on that central value, achievable in past measurements of quenching in inorganic scintillators. With this, COH-CryoCsI-1 would have a $\approx500$~eV$_\mathrm{nr}$ threshold for nuclear recoils.  Though this is preliminary, COHERENT plans a more rigorous measurement of the quenching factor also looking at variation with temperature.

It is also worth noting that increased light yield, beyond lowering threshold, will also improve both detector timing and energy resolution.  Recoil time for a CEvNS interaction is determined from the first observed PE pulse in a reconstructed waveform.  For a scintillator, this resolution is the scintillation time constant divided by the number of PE pulses observed.  With its high light yield, COH-CryoCsI-1 will have precise timing resolution, allowing excellent separation of prompt and delayed CEvNS.  Similarly, the energy resolution scales as $1/\sqrt{N}$ if photon counting dominates the resolution.  This will be advantageous for extracting physics that distorts the $Q^2$ dependence of the CEvNS cross section, such as measuring nuclear form factors and testing the weak charge distribution.  

\section{Searching for low-mass mediators of new forces}

The earliest CEvNS results~\cite{COHERENT:2017ipa,COHERENT:2020iec,XENON:2020gfr,COHERENT:2021xmm,CONUS:2021dwh} have successfully demonstrated CEvNS as a powerful probe of neutrino non-standard interactions (NSIs)~\cite{Miranda:2020tif,Denton:2020hop,PhysRevD.104.015019,PhysRevD.105.033001}.  NSIs would be a natural consequence of a new force that couples feebly to standard-model particles and give rise to flavor-dependent anomalous couplings between neutrinos and quarks.  The possible NSIs are usually described by a general effective Lagrangian parameterized by the tensor of couplings $\varepsilon_{\alpha\beta}^{q}$, where $\alpha$ and $\beta$ are initial and final neutrino flavors $(e, \mu, \tau)$ and $q$ is quark flavor $(u,d)$~\cite{Miranda:2004nb,Friedland:2004pp,Proceedings:2019qno}.  The presence of neutrino NSIs would alter neutrino flavor transitions in matter, leading to ambiguities in NSI and neutrino-mixing parameter space~\cite{Gonzalez-Garcia:2013usa,PhysRevD.94.055005,Coloma:2017egw,Coloma:2017ncl}.

Within the standard model, neutrinos propagating through matter experience an increased effective mass due to low-$Q^2$, forward $\nu-e$ elastic scattering.  This is a purely neutral current (NC) interaction for $\nu_\mu$ and $\nu_\tau$ flavors, but for $\nu_e$ neutrinos, both NC and charged current (CC) diagrams contribute, leading to a higher scattering potential.  Thus, the $\nu_e$ flavor propagates with a different effective mass.  This contributes an additional term to the vacuum oscillation Hamiltonian of
\begin{equation}
    H^\prime = A
    \begin{pmatrix}
    1+\varepsilon_{ee} & \varepsilon_{e\mu} & \varepsilon_{e\tau} \\
    \varepsilon_{\mu e} & \varepsilon_{\mu\mu} & \varepsilon_{\mu\tau} \\
    \varepsilon_{\tau e} & \varepsilon_{\tau\mu} & \varepsilon_{\tau\tau}
    \end{pmatrix},
\label{eqn:matterEffects}
\end{equation}
in the flavor basis where $A=\sqrt{8}G_FN_eE_\nu$, $G_F$ is the Fermi constant, $N_e$ is the density of electrons in matter, $E_\nu$ is the neutrino energy, and $\varepsilon_{\alpha\beta}$ have been summed over $u$, $d$ couplings weighted by the relative densities of quarks and electrons in matter.  By assuming $\varepsilon_{ee}=-2$ with other couplings $0$ and adjusting the neutrino mixing parameters as in~\cite{Coloma:2017ncl}, the Hamiltonian is transformed as $H\rightarrow-H^\ast$.  This scenario preserves oscillation dynamics and is not testable with oscillation experiments alone~\cite{Denton:2022nol}.  Futher, since oscillations do not depend on the absolute mass scale, adding a multiple of the identity matrix to $H^\prime$ also has no effect.  Thus, there is a linear space of NSI parameters -- $\varepsilon_{ee}=-2+x$, $\varepsilon_{\mu\mu}=\varepsilon_{\tau\tau}=x$ -- that would imply dramatically different oscillation parameters from those typically quoted such as $\Delta m^2_{32}\rightarrow-\Delta m^2_{32}$ and $\delta_{CP}\rightarrow\pi-\delta_{CP}$~\cite{NOvA:2021nfi,T2K:2023smv}.  These two solutions described by the absence or presence of NSIs are termed the large mixing angle (LMA) and LMA-Dark solutions, respectively.  Data from scattering experiments are required to resolve this degeneracy~\cite{Denton:2022nol}.  There must be NSIs for either $\nu_e$ $(\varepsilon_{ee}=-2)$, or $\nu_\mu$ and $\nu_\tau$ $(\varepsilon_{\mu\mu}=\varepsilon_{\tau\tau}=2)$, for LMA-Dark to hold.  COHERENT detectors, in the multi-flavor $\pi$DAR neutrino flux at the SNS, can test both $\varepsilon_{ee}$ and $\varepsilon_{\mu\mu}$.  Thus COHERENT can address the LMA vs LMA-Dark question.

In effective field theory, early results from COHERENT's first CsI detector strongly disfavor NSI couplings required to satisfy the LMA-Dark scenario.  The COH-CryoCsI-1 and future upcoming 18-kg germanium (COH-Ge-1) and 750-kg argon (COH-Ar-740) CEvNS detectors will all improve on constraints of these NSI couplings~\cite{Akimov:2022oyb}.  However, for mediator masses below the momentum transfer at the CsI threshold, $\sqrt{Q^2}\approx$~40~MeV, this approach of treating NSI effects at simple scale factors becomes invalid and LMA-Dark remains viable.  Fortunately, the parameter space is bounded; a mediator lighter than 3.1~MeV would affect big bang nucleosynthesis and is ruled out by cosmology~\cite{Sabti_2021}.  Thus, there is an opportunity for upcoming CEvNS detectors to decisively resolve the LMA vs LMA-Dark question by improving the thresholds of CEvNS, bridging the gap between current constraints from cosmology and neutrino scattering.  With an $\approx$~$80$~eV$_\mathrm{ee}$ threshold, multiple neutrino flavors accessible in the SNS flux, and the detector's fast timing, COH-CryoCsI-1 is unique among COHERENT's future detectors in satisfying all these requirements.  It can conclusively clarify the ambiguity between the NSI and neutrino-mixing landscapes ahead of precision oscillation data with DUNE, T2HK, and JUNO~\cite{Abi_2020,DUNE:2020ypp,Hyper-Kamiokande:2016srs,Zhang:2021adu}.


A mediator of a new force with the same strength as the weak force would yield neutrino NSI effects of order $\varepsilon\sim1$.  There may be NSI effects of a similar strength to the standard-model weak couplings, but such small effects would be incredibly challenging to detect in strong or electromagnetic interactions.  Measurements with unprecedented precision, however, may detect the subtle influence of the new mediator.  Perhaps the most famous example is the $g-2$ measurement of the muon anomalous magnetic moment at BNL~\cite{PhysRevD.73.072003} and FNAL~\cite{PhysRevLett.126.141801,Muong-2:2023cdq} which has observed a notable discrepancy with standard-model calculations~\cite{Aoyama:2020ynm}.  This discrepancy is often attributed to a dark photon that interferes with the standard-model photon~\cite{Athron:2021iuf} and is also predicted to interfere in neutrino scattering amplitudes.  Consequently, searches for neutrino NSIs at CEvNS experiments are an additional and direct probe of new mediators that may explain $g-2$ by testing flavor-dependent NSI effects in neutrino scattering.  

The standard-model differential CEvNS cross section can be approximately written as 
\begin{equation}
    \frac{d\sigma}{dE_r} = \frac{Q_W^2}{2\pi}\left[1-\frac{2m_NE_r}{E_\nu^2}\right]\left|F(Q^2)\right|^2
    \label{eqn:CEvNSXSec}
\end{equation}
where $E_r$ is the nuclear recoil energy, $m_N$ is the nuclear mass, $E_\nu$ is the incoming neutrino energy, and $Q_W = g_pZ + g_nN$ is the nuclear weak charge with $Z$ and $N$ the proton and neutron numbers of the nucleus.  In the standard model, $g_p^\mathrm{SM}=1/2-2\sin^2\theta_{W}$ and $g_n^\mathrm{SM} = -1/2$.  If NSIs from a heavy vector mediator are included and we assume neutrinos couple equally to $u$ and $d$ quarks, these couplings are adjusted for neutrino flavor $\alpha$ as $g_p\rightarrow g_p^\mathrm{SM}+3\varepsilon_{\alpha\alpha}$ and $g_n\rightarrow g_n^\mathrm{SM}+3\varepsilon_{\alpha\alpha}$.  In the case that the mediator mass is not $\gg \sqrt{Q^2}$, the $\varepsilon_{\alpha\alpha}$ become functions of $Q^2$:
\begin{equation}
    \varepsilon_{\alpha\alpha}(Q^2) = \frac{g_qg_\alpha}{\sqrt{8}G_F}\frac{1}{m_V^2+Q^2}
\end{equation}
where $g_q$ and $g_\alpha$ are the $q$ and $\nu_\alpha$ charges under the new forces and $m_V$ is the mediator mass.  Thus, NSIs affect the CEvNS shape as well as the rate.

\begin{figure}
  \includegraphics[width=0.48\textwidth]{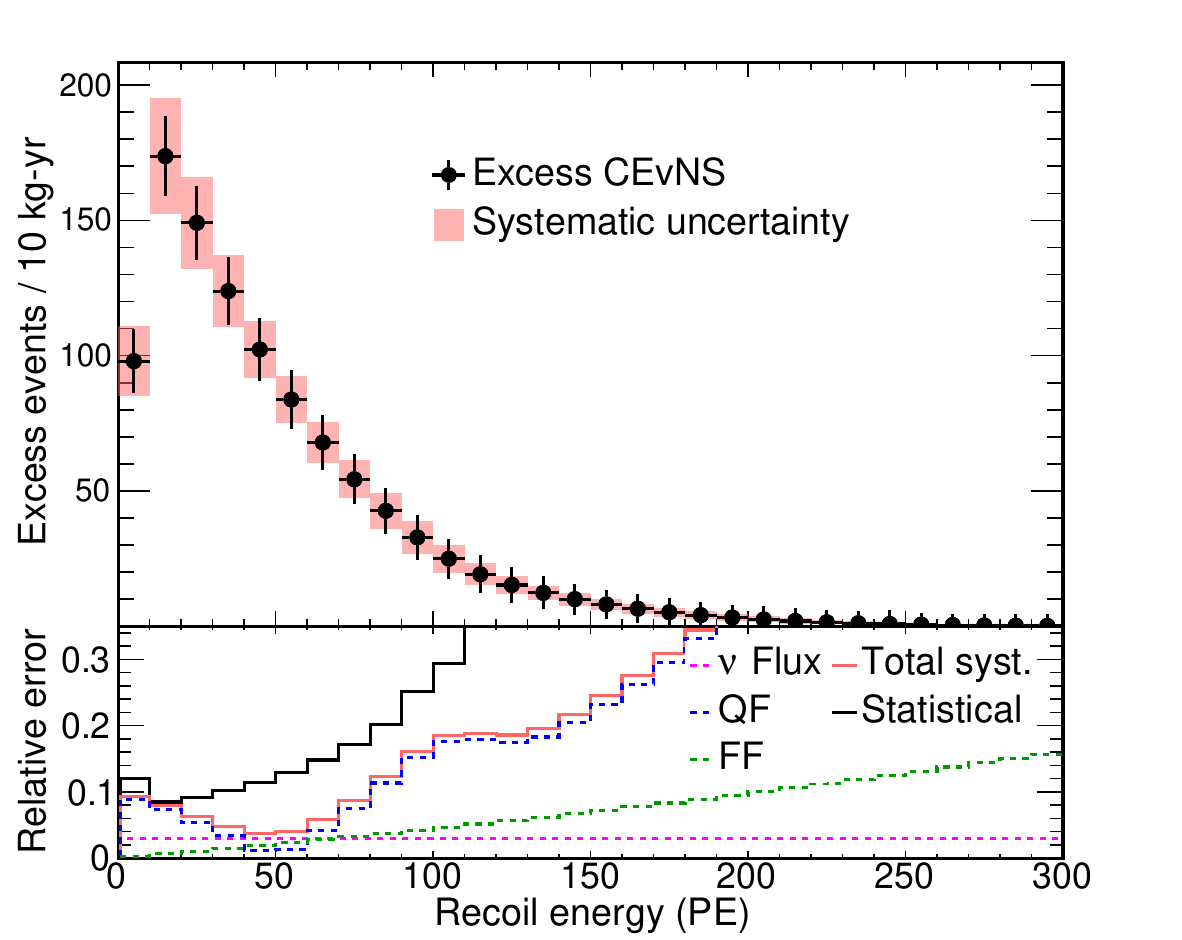}
  \caption{The expected CEvNS rate with statistical errors after background subtraction and systematic errors (red shaded region). Below, we show the decomposition of the systematic uncertainty into neutrino flux, quenching (QF), and form-factor uncertainties (FF).}
  \label{fig:systCocktail}
\end{figure}

Since light-mediator NSIs would affect the CEvNS recoil distribution, systematic uncertainties that distort the shape must be properly accounted for.  For this preliminary estimate of the detector's reach, we considered three sources of systematic uncertainty: neutrino flux, quenching, and nuclear form factor, which are summarized with statistical errors in Fig.~\ref{fig:systCocktail}.  The neutrino flux uncertainty, currently $10\%$ from a comparison of simulation to hadron production data~\cite{PhysRevD.106.032003}, was taken as $3\%$ which is achievable with COHERENT's flux calibration efforts~\cite{Akimov_2021}.  For other uncertainties, we accounted for expected spectral distortions on CEvNS selected by our cuts.

We assume that the quenching of scintillation from nuclear recoils is a flat $(15\pm1.5)\%$ as a preliminary estimate; analysis of quenching data is ongoing.  Quenching dramatically change the analysis threshold COH-CryoCsI-1 can achieve and gives a large uncertainty at low recoil energies where light-mediator NSI effects are most prominent.  As this affects the CEvNS sample in the most sensitive kinematic region, we plan additional quenching measurements that extend lower in energy, near the detector threshold.  

Neutrino timing will be important for separating spectral distortions arising from physics and systematic effects at low $Q^2$.  Because prompt $\nu_\mu$ and delayed $\nu_e/\bar{\nu}_\mu$ neutrino fluxes are separated in time, it will be possible to detect different couplings to $\nu_e$ and $\nu_\mu$ neutrino flavors even with large uncertainties at the lowest recoil energies.  This is a key advantage of COH-CryoCsI-1 over semiconductor detectors such as COH-Ge-1, which achieves similarly low recoil thresholds but with worse timing resolution.

The nuclear form factor gives the degree to which the extent of the weak nuclear charge is finite rather than point-like.  At large $Q^2$, the diffuse charge distribution leads to incoherence, reducing the CEvNS cross section.  We used the Klein-Nystrand form factor~\cite{Klein:1999qj} which is parameterized by the neutron radius, $R_n$.  We took a 5$\%$ uncertainty in $R_n$.  This has very little effect on CEvNS events near threshold, but increases with recoil energy.  At 20~keV$_\mathrm{nr}$, this uncertainty in $R_n$ translates to a $\pm5\%$ uncertainty in the CEvNS rate.  Conversely, this suppression at high $Q^2$ allows CEvNS to constrain the weak charge radius of the nucleus, which we describe in Sect.~\ref{sect:Rn}.

There is strong complementarity between measurements at reactors and accelerators.  The neutrino flux at the former is entirely $\bar{\nu}_e$ while the flux at the latter contains $\nu_\mu$, $\nu_e$, and $\bar{\nu}_\mu$ flavors.  As such, first we considered a scenario where reactor constraints have unambiguously determined the coupling $\varepsilon_{ee}=0$ and later considered a scenario with no reactor input, treating both $\varepsilon_{ee}$ and $\varepsilon_{\mu\mu}$ to be completely free parameters.  

We have estimated the sensitivity of a three-calendar-year run of a 10~kg CryoCsI detector at the SNS to light mediator neutrino-quark NSIs given the expected background and signal sample determined in Sect.~\ref{sect:Detector} with a particular emphasis on resolving the neutrino-oscillation ambiguity.  We performed a 2D log-likelihood fit using Asimov fake data~\cite{Cowan:2010js} generated with no NSI effects.  The fake data were binned in both recoil energy and time to detect the expected $Q^2$-dependent distortions and separate $\nu_e$ and $\nu_\mu$ flavors.  Penalty terms for the three systematic uncertainties (neutrino flux, quenching, and form-factor suppression) were included in the likelihood.  For each set of true model parameters, $\left(m_V,\varepsilon_{ee},\varepsilon_{\mu\mu}\right)$, we calculated a $-2\Delta\log\mathcal{L}$ by profiling over all nuisance parameters.  From this, we calculated the $2\sigma$ exclusion curves we would draw if no new physics were detected.  We assumed Gaussian statistics with 2D critical $\Delta\chi^2$ values and allowed only two physics parameters to vary at a time.

\begin{figure}
  \includegraphics[width=0.48\textwidth]{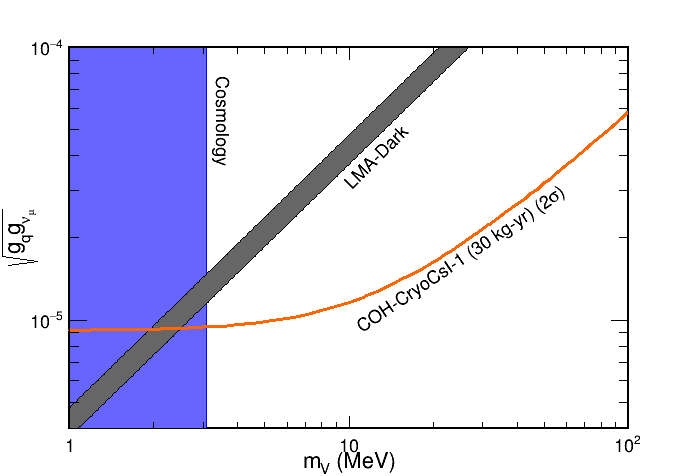}
  \caption{A summary of LMA-Dark viable NSI parameter space (gray band) compared to current constraints from cosmology (blue vertical region).  COH-CryoCsI-1 will disfavor couplings above the orange curve, assuming the LMA solution.  CEvNS data from nuclear reactors is implicitly assumed to fully test the $\nu_e$ NSI coupling.}
  \label{fig:LowMassOscContour}
\end{figure}

\textbf{Case 1: NSI constraints available from reactor experiments}.  First, assuming that $\varepsilon_{ee}$ is known to be 0 from reactor data, we tested the $\nu_\mu$ coupling of the NSIs as a function of mediator mass.  The expected sensitivity of COH-CryoCsI-1 is shown in Fig.~\ref{fig:LowMassOscContour} compared with additional constraints from cosmology.  The diagonal gray band gives the LMA-Dark parameter space, $\varepsilon_{\mu\mu}\approx2$, that is allowed by oscillation data.  At the lowest mediator masses where LMA-Dark remains viable, we could detect $\nu_\mu$-coupled NSIs at over $3\sigma$.

\begin{figure*}
  \includegraphics[width=0.48\textwidth]{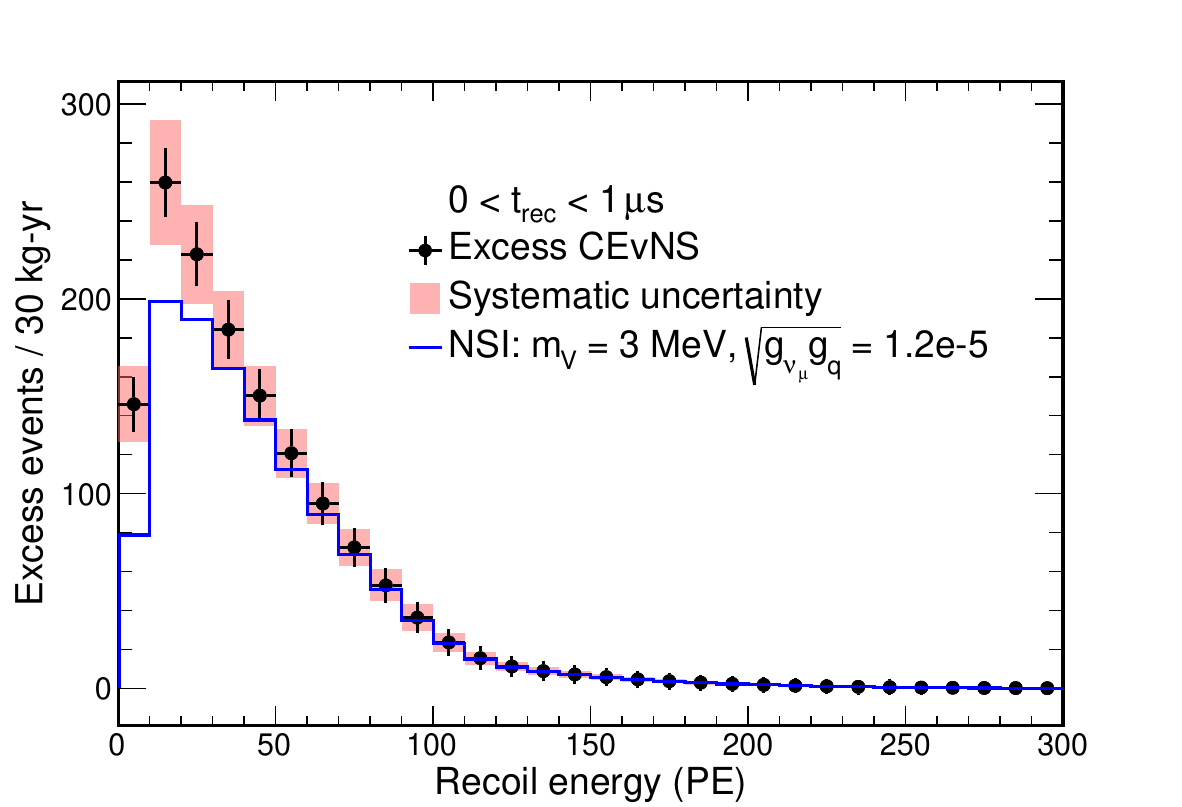}
  \includegraphics[width=0.48\textwidth]{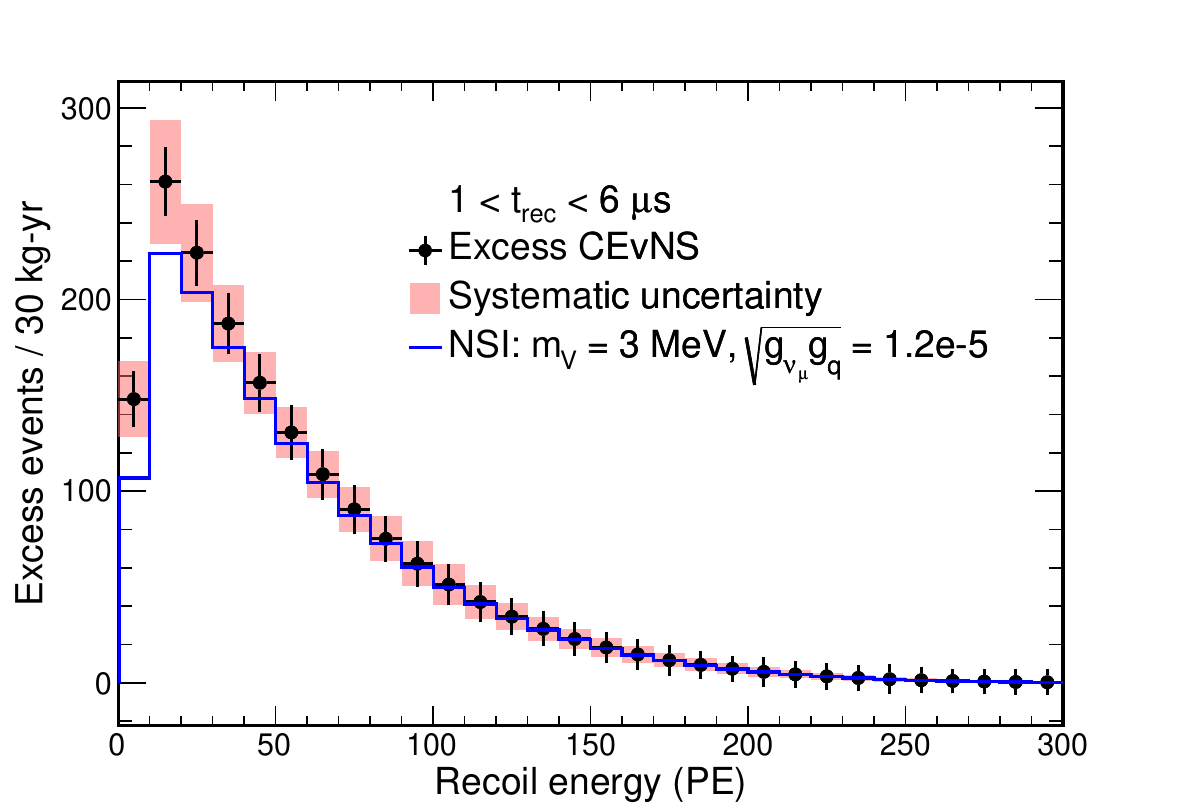}
  \includegraphics[width=0.48\textwidth]{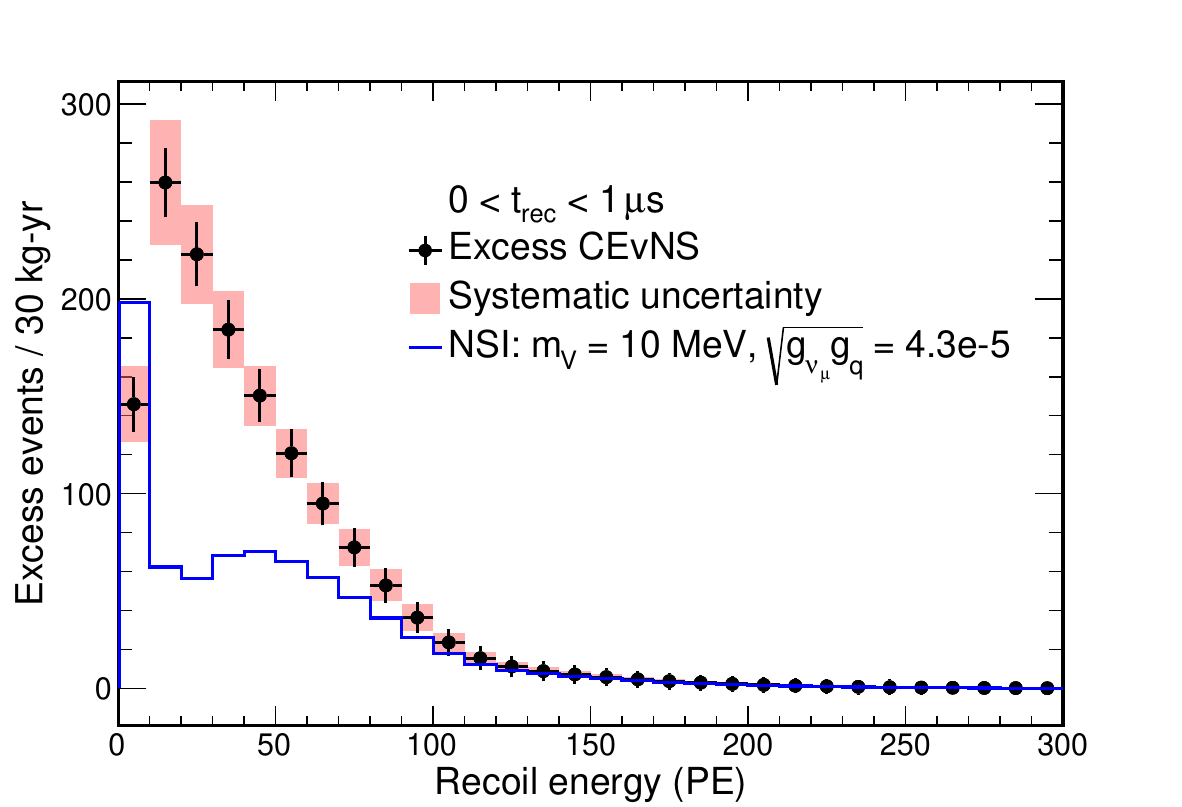}
  \includegraphics[width=0.48\textwidth]{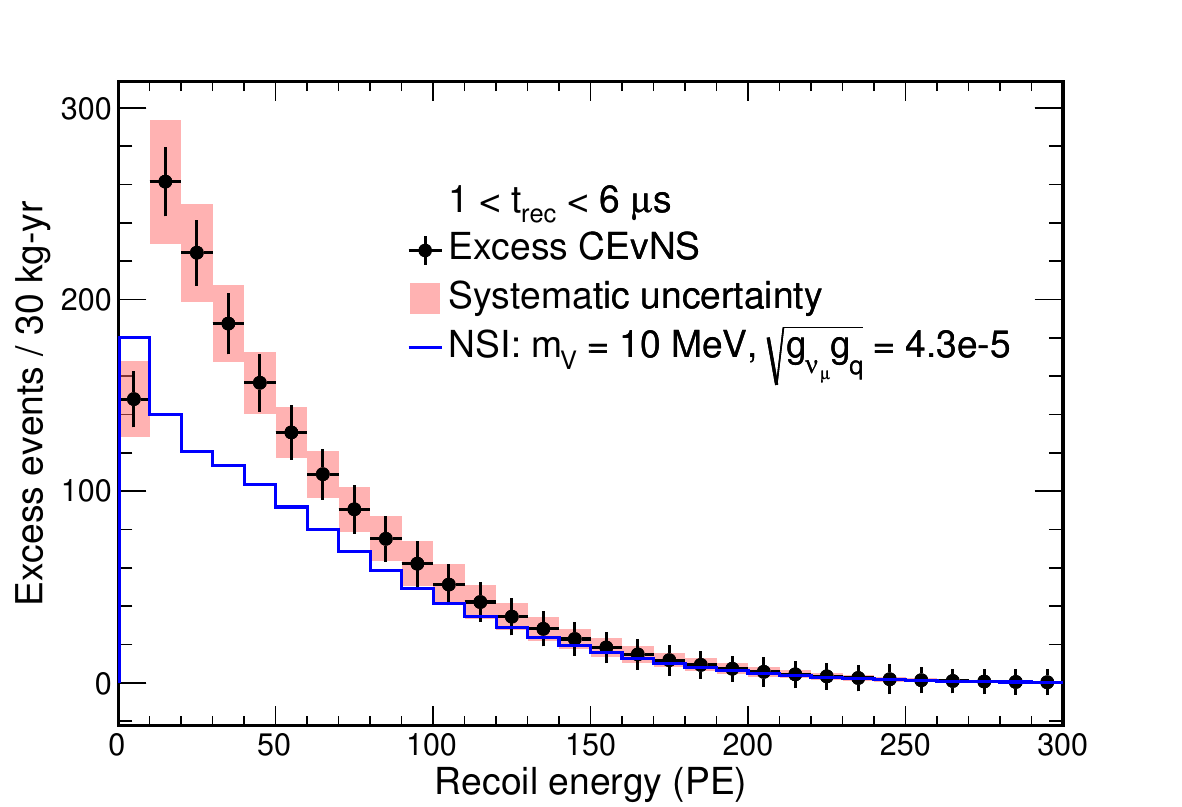}
  \caption{Background-subtracted CEvNS spectra expected in COH-CryoCsI-1 in the prompt (left, $t_\text{rec}<1$~$\mu$s) and delayed (right, $t_\text{rec}\geq1$~$\mu$s) timing regions of interest.  For each case, we compare the expected standard-model prediction with statistical and systematic uncertainties to the central-value prediction in the presence of a light mediator coupling neutrinos and quarks.  We chose two mediator masses: 3~MeV (top) and 10~MeV (bottom).}
  \label{fig:LowMassSpectra}
\end{figure*}

For masses near the cosmological limit, the recoil energy shape is only distorted near the detector threshold.  To show how this compares to the large uncertainty on the event rate from quenching in this region, we show the standard-model prediction with systematic error band and the distorted spectrum expected for a LMA-Dark scenario with mediator mass fixed at $3$ and $10$~MeV in Fig.~\ref{fig:LowMassSpectra}.  Here, we assumed that the NSIs giving rise to the LMA-Dark solution are entirely in the $\nu_\mu$ and $\nu_\tau$ flavors.  The sample was subdivided further to those events with recoil times $<1\mu s$ and $\geq 1\mu s$, which select subsamples of CEvNS from nearly pure $\nu_\mu$ and from a mix of $\nu_e$ and $\bar{\nu}_\mu$ flavors, respectively.  Thus, the NSI effects are more pronounced in the $<1\mu s$ sample.  In the $m_V=3$~MeV case, the NSIs suppress the prompt (delayed) event count by 50$\%$ (30$\%$) for PE~$<10$ so that even a large uncertainty in the quenching can be distinguished from the NSI physics to a certain degree.  For the LMA-Dark parameter space with $m_V>10$~MeV, the spectrum is very dramatically altered and COH-CryoCsI-1 can distinguish between the LMA and LMA-Dark hypotheses at very high significance.

\begin{figure}
  \includegraphics[width=0.48\textwidth]{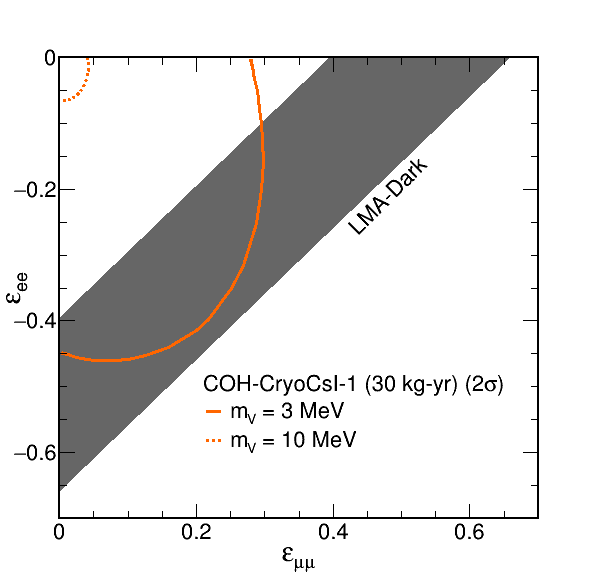}
  \caption{Capacity for COH-CryoCsI-1 to test the NSI parameter space consistent with LMA-Dark by neutrino coupling, both $\nu_e$ and $\nu_\mu/\nu_\tau$ flavors, to quarks, assuming the LMA solution.  No additional NSI constraint from reactor CEvNS experiments is assumed. We compute the sensitivity both at a mediator mass of 3~MeV, at the cosmological limit, and at a higher mass, 10~MeV.}
  \label{fig:LowMassOscCouplings}
\end{figure}

\textbf{Case 2: no input from reactor experiments}.  We also considered the scenario where both $\varepsilon_{ee}$ and $\varepsilon_{\mu\mu}$ are unconstrained parameters, temporarily neglecting any constraints from CEvNS experiments at nuclear reactors.  The same fitting procedure was applied to test the COH-CryoCsI-1 sensitivity in this scenario, again after 3 years of SNS running, and shown in Fig.~\ref{fig:LowMassOscCouplings}.  Resulting 2$\sigma$ contours are shown as a function of the LMA-Dark NSI parameter space.  Though COH-CryoCsI-1 could distinguish between LMA and LMA-Dark in the $\nu_\mu$ flavor assuming $\varepsilon_{ee}=0$, the detector cannot test all parameter space for non-zero values of the $\nu_e$ coupling for $m_V\approx3$~MeV.  In this sense, innovations in low-threshold CEvNS detectors at reactors~\cite{Bonet:2023kob,NUCLEUS:2019igx} are as essential as COH-CryoCsI-1 to determining the neutrino mixing landscape.  Reactors have sensitivity to $\varepsilon_{ee}$ from the large $\bar{\nu}_e$ flux produced during fission, placing a horizontal contour in Fig.~\ref{fig:LowMassOscCouplings}.  To fully understand the interplay of light-mediator NSIs and neutrino oscillations, reactors would need to test $\varepsilon_{ee}<0.1$ so that, when combined with COH-Cryo-CsI-1, the LMA-Dark solution can be fully explored at $>2\sigma$.  


Apart from applications to neutrino-oscillation experiments, COH-CryoCsI-1 will also directly test a new-force explanation of the anomalous muon magnetic moment observed in $g-2$.  Though the size of the anomaly determines the charge of the $\mu$ under this new force, couplings to other standard-model fermions could generally be free.  A new force with universal couplings is strongly disfavored from electron scattering experiments which show agreement with the standard-model couplings even at low $Q^2$~\cite{Athron:2021iuf}.

As an interesting example, the anomaly-free $L_\mu-L_\tau$ symmetry may arise from a gauge boson, resulting in a $U(1)$ dark-photon, $V$, extension to the standard model which can explain $g-2$~\cite{Baek:2001kca,Ma:2001md}.  Such a simple solution would have profound implications if directly observed and would naturally explain dark matter~\cite{Biswas:2016yan,Biswas:2016yjr,Patra:2016shz} and the neutrino masses~\cite{Heeck:2011wj,Baek:2015mna,Biswas:2016yan}.  However, as only $\mu$, $\tau$, and their corresponding neutrinos would be charged under such a force, this is among the most elusive dark photon explanations of $g-2$.  However, COHERENT can test the muon coupling due to the $\nu_\mu/\bar{\nu}_\mu$ flux at the SNS~\cite{Banerjee:2021laz,AtzoriCorona:2022moj}.  Constraints from first-light detectors fall just short of other constraints, but future CEvNS detectors with larger sample sizes and lower thresholds can exhaustively test the $g-2$ favored parameter space for the $L_\mu-L_\tau$ model~\cite{AtzoriCorona:2022moj}.  

To tree level, only $\mu$, $\tau$ generations of leptons are charged and this new force has no effect on the CEvNS cross section.  However, the gauge boson can mix with the photon by virtual $\mu$ or $\tau$ bubble which then couples $\nu_\mu$ neutrinos with the proton number of a nucleus.  This modifies the weak charge of proton coupling in $Q_W$ of the nucleus as 
\begin{equation}
  g_p\rightarrow g_p^\mathrm{SM}+\frac{g_V^2}{\sqrt{18}\pi G_F}\alpha_\mathrm{EM}\log{\left(\frac{m_\tau}{m_\mu}\right)^2}\left(\frac{1}{m_V^2+Q^2}\right)
\end{equation}
where $g_V$ is the charge of the new force for $\mu$ and $\tau$ lepton generations.  In this sense, the $L_\mu-L_\tau$ model induces $Q^2$-dependent effects analogous to low-mediator NSIs for the $\nu_\mu/\bar{\nu}_\mu$ fluxes present at the SNS.  The same framework developed for testing LMA-Dark can thus be directly applied to this problem.  Notably, this affects the proton coupling, as opposed to CEvNS which preferentially couples to neutrons.  Thus, a positive detection in COH-CryoCsI-1 may be disentangled from mismodeling of the standard-model background by comparing CEvNS on multiple targets with different $N/Z$ ratios.

\begin{figure}
  \includegraphics[width=0.48\textwidth]{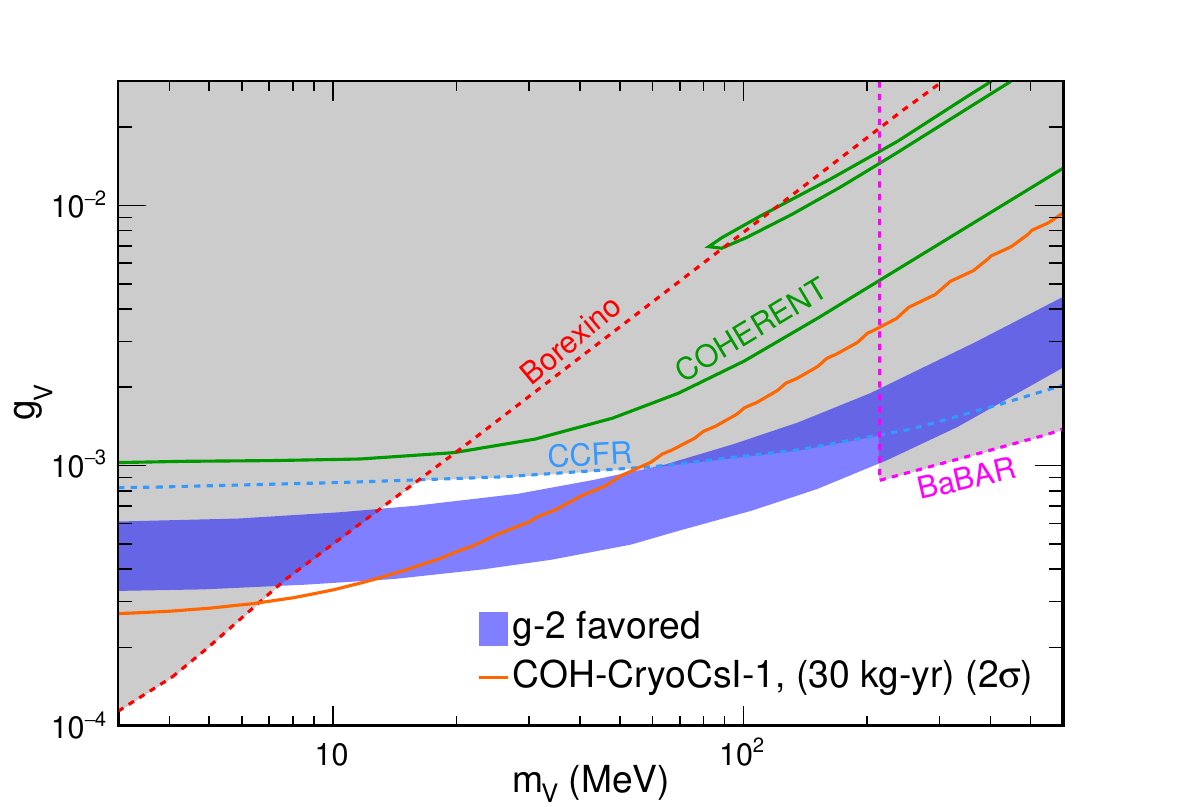}
  \caption{Sensitivity of COH-CryoCsI-1 to a $L_\mu-L_\tau$ mediator compared to current constraints from CEvNS (solid lines) and other experiments (dashed lines).  Such a model would resolve the reported $g-2$ anomaly in the parameter space given by the blue shaded region.}
  \label{fig:LmuLtau}
\end{figure}

Thus, we determined the $L_\mu-L_\tau$ parameter space that COH-CryoCsI-1 will be sensitive to with a 2D fake data sensitivity fit.  As scattering of $\nu_e$ is unaffected, timing is similarly important.  The sensitivity after three years is shown in Fig.~\ref{fig:LmuLtau}.  The model is only viable for mediator masses between roughly 10~MeV and 200~MeV, below which Borexino~\cite{Bellini:2011rx} and above which CCFR~\cite{PhysRevLett.66.3117,Altmannshofer:2014pba}, BaBAR~\cite{BaBar:2016sci}, and CMS~\cite{CMS:2018yxg} (sensitive at $m_V>4$~GeV outside the region of interest of the plot) can rule out any parameter space consistent with $g-2$.  Similar to the LMA vs LMA-Dark question, this leaves a bounded parameter space for future experiments to explore.  Results from the COH-CryoCsI-1 detector will test about half of the remaining $L_\mu-L_\tau$ parameter space which would explain the $g-2$ anomaly.  The remainder could be explored~\cite{AtzoriCorona:2022moj} with an upgraded ton-scale cryogenic CsI CEvNS detector placed at the SNS second target station~\cite{Asaadi:2022ojm}.

\section{Potential to discover hidden-sector particles}

The SNS is a world-leading neutron and neutrino production facility, with operations planned to increase to 2.0~MW.  This intensity also makes the SNS an excellent beam-dump facility.  Hidden-sector particles may be produced abundantly through anomalous decays of meson such as $\pi^0/\eta^0$ as a consequence of the $\sim$~2~$\times10^{23}$ protons on target delivered each year, and may scatter or decay in COHERENT detectors, leaving a visible signature which can be observed over the CEvNS excess.  CEvNS detectors show novel sensitivity to vector~\cite{PhysRevD.70.023514,BOEHM2004219,Pospelov:2007mp,PhysRevD.84.075020,deNiverville:2015mwa,Dutta:2020vop,Battaglieri:2017aum} and axion-like particle (ALP) portals to the hidden sector~\cite{PhysRevD.75.052004,AristizabalSierra:2020rom,PhysRevLett.126.201801,CCM:2021jmk}.  The study of ALP detection at the SNS is underway, and here we describe sensitivity to a DM model that features kinetic mixing between the photon and a vector portal particle already studied by COHERENT in the original CsI detector~\cite{COHERENT:2019kwz,COHERENT:2021pvd}.  

\begin{figure}
  \includegraphics[width=0.48\textwidth]{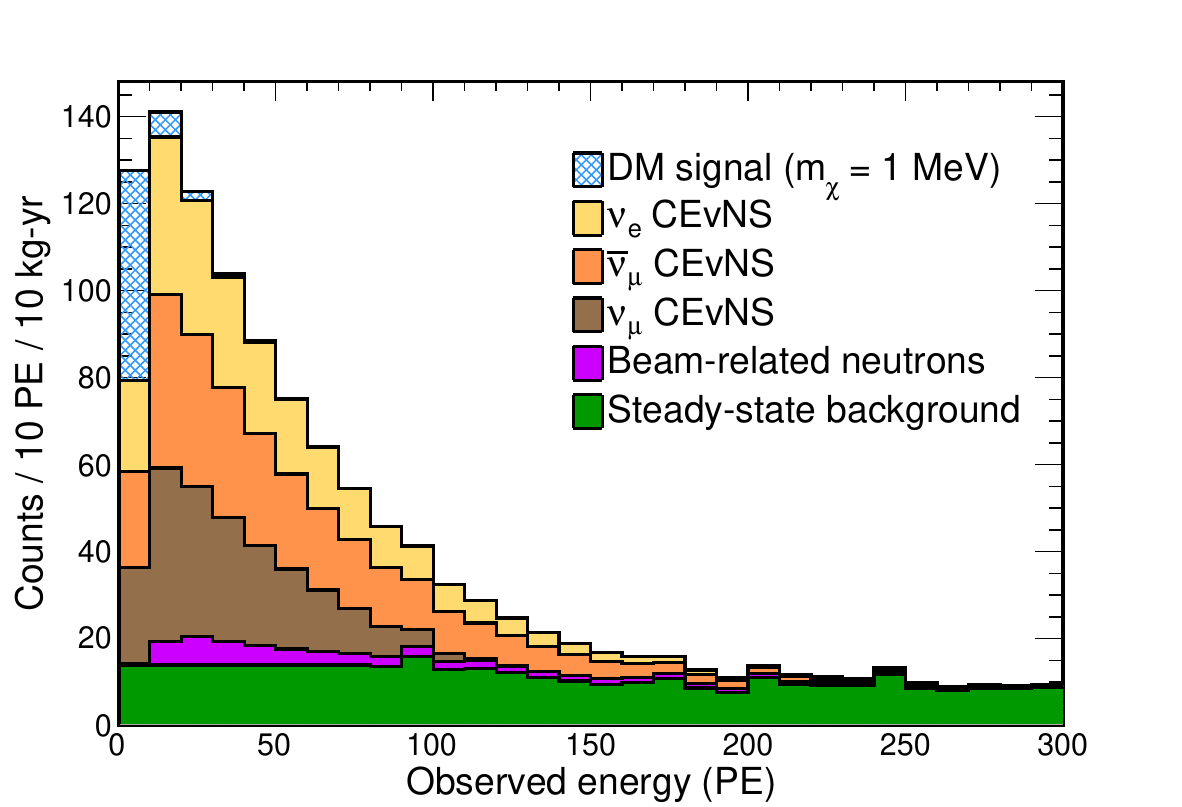}
  \includegraphics[width=0.48\textwidth]{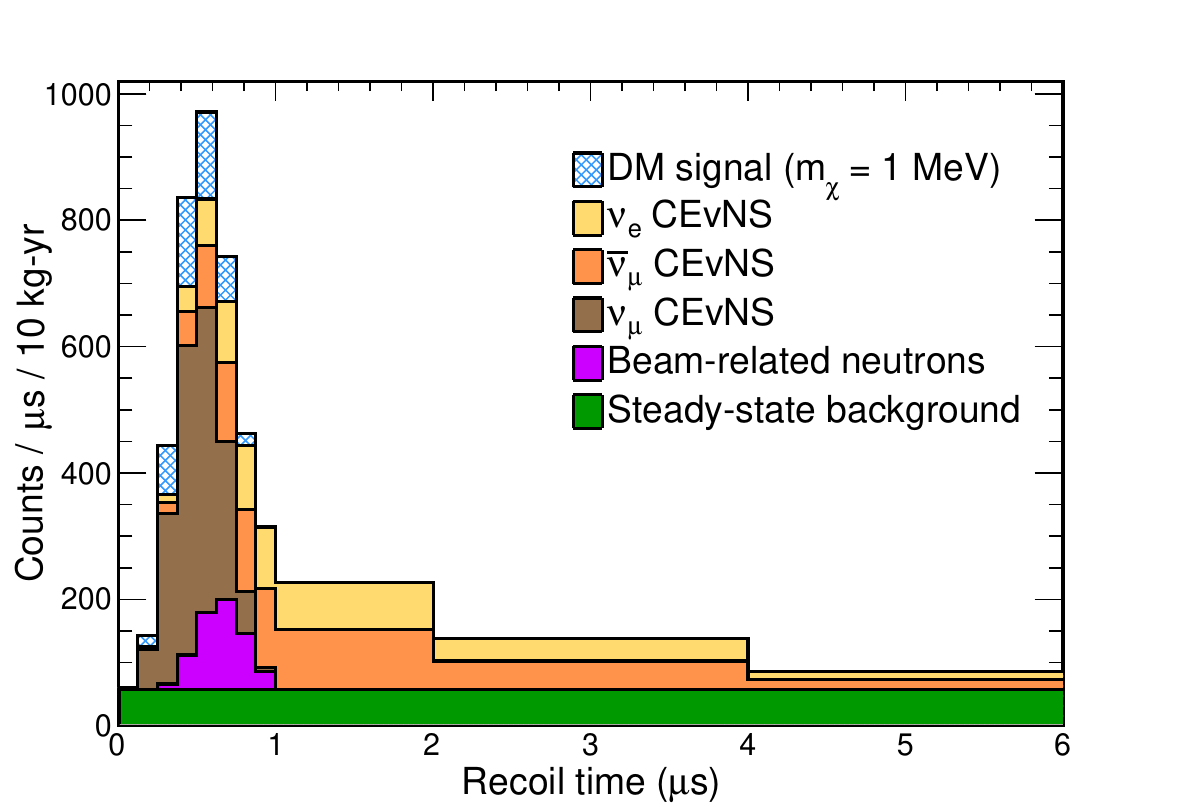}
  \caption{Expected spectra of selected events, including dark matter, CEvNS, and backgrounds, after one year of SNS running.  A mass of 1~MeV is assumed for the dark matter.  Dark matter (blue hatched) would manifest as an additional scattering process at early times.  Later CEvNS from $\nu_e$ and $\bar{\nu}_\mu$ would improve systematic uncertainties on the detector response and neutrino flux to inform the background in the dark matter region of interest, $t_\mathrm{rec}<1\mu$s.}
  \label{fig:CryoCsIDMSpectra}
\end{figure}

As in the low-mediator NSI case, the low threshold and favorable timing resolution of the cryogenic scintillator technology give this detector a much higher efficiency.  This is particularly true for dark matter with mass $m_\chi<10$~MeV which produces a softer energy spectrum.  The expected energy and time distributions of DM-induced nuclear recoils are shown in Fig.~\ref{fig:CryoCsIDMSpectra} for the lowest DM mass we considered, $m_\chi=1$~MeV.  At this mass, the efficiency was estimated at 11$\%$ in the COH-CryoCsI-1 detector compared to $0.09\%$ in COHERENT's first CsI detector, an improvement of over two orders of magnitude.

\begin{figure}
  \includegraphics[width=0.48\textwidth]{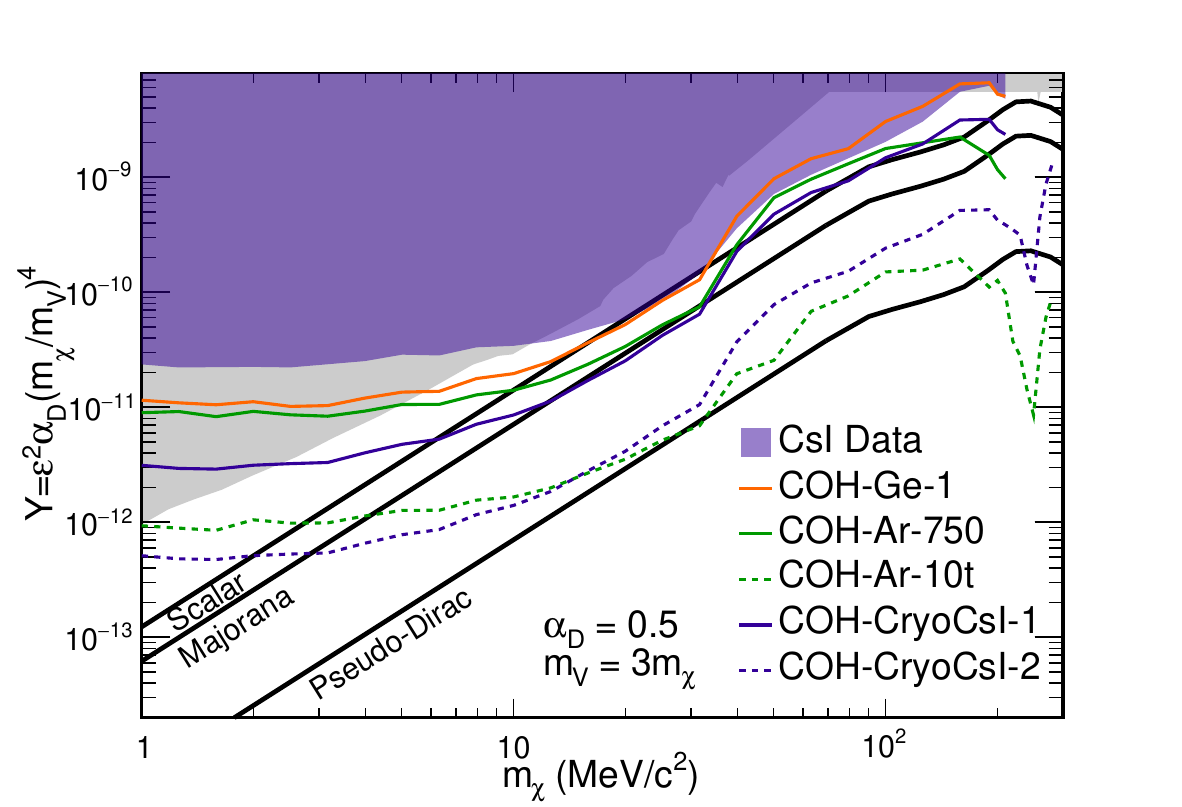}
  \caption{Calculated sensitivities to new parameter space testable with future CEvNS detectors at the SNS.  Cryogenic CsI detectors perform well, particularly at low dark-matter masses.  The COH-CryoCsI-2 detector is an upgraded, 700~kg detector at the SNS using the same technology.}
  \label{fig:DMSens}
\end{figure}

The timing distribution for DM recoils is very favorable at the SNS.  Any DM produced would arrive coincident with the prompt $\nu_\mu$ flux.  As shown in Fig.~\ref{fig:CryoCsIDMSpectra}, there is a large number of delayed CEvNS scatters which can be used to constrain the detector response uncertainties and improve the prediction of the signal neutrino spectrum~\cite{COHERENT:2019kwz}.  Due to this uncertainty reduction, CEvNS experiments do not become limited by systematic uncertainties at exposures possible at the SNS.  In~\cite{COHERENT:2021pvd}, the dark-matter sensitivity has been calculated for two detectors: COH-CryoCsI-1 and later a 700-kg undoped, cryogenic CsI detector that would be housed at the SNS second target station.  Expected contours are shown in Fig.~\ref{fig:DMSens}.  With its low threshold, even the 10-kg COH-CryoCsI-1 detector would have sensitivity to new DM parameter space beyond any other detector to be commissioned in Neutrino Alley for $m_\chi<20$~MeV.  This will test the theoretically motivated relic abundance lines for both scalar and Majorana fermion DM over a significant portion of the surveyed parameter space.

\section{Searching for sterile neutrinos with CEvNS disappearance}

Neutrino-oscillation experiments are constantly improving understanding of neutrino mixing.  Interestingly, some observed oscillation signatures are inconsistent with the three-flavor paradigm and could be explained by one or more additional sterile neutrino states.  The LSND experiment~\cite{LSND:2001aii} first reported such a signature using accelerator neutrinos, followed by MiniBooNE~\cite{MiniBooNE:2020pnu}.  Later, similar anomalies were detected in gallium experiments~\cite{SAGE:1998fvr,Abdurashitov:2005tb,Kaether:2010ag,PhysRevLett.128.232501,Elliott:2023cvh}.  Evidence for sterile oscillations has been found in reactor-based experiments~\cite{Mention:2011rk} but may instead be a consequence of poor understanding of fission products inside reactors~\cite{PhysRevLett.130.021801}.  These results can be explained with a sterile neutrino state with a mass splitting $\Delta m^2_{41}\approx 1.7$~eV$^2$ determined from a global fit~\cite{Gariazzo:2017fdh}.  However, many experiments running similar searches have found results that are inconsistent with a sterile neutrino to a large significance~\cite{KARMEN:2002zcm,MINOS:2016viw,IceCube:2016rnb,MicroBooNE:2021rmx}.  Many of these experiments use neutrino scattering at $\sim1$~GeV where neutrino interaction uncertainties complicate the interpretation of many experimental results.  CEvNS, however, is very cleanly calculated in the standard model so that precision CEvNS datasets evade the complicated interaction modeling.

The COHERENT experiment has deployed CEvNS detectors at multiple baselines between 19 and 28~m.  The neutrino energies, in the 10s of MeV, are ideal for testing mass splittings of $\approx2$~eV$^2$, very near the best fit.  Since CEvNS is a NC process, it is insensitive to three-flavor oscillations.  But, a sterile neutrino does not participate in the weak force, so oscillations from active to sterile states would be observable as a reduction of the CEvNS rate~\cite{Blanco:2019vyp,Bisset:2023oxt}, depending on the baseline and neutrino energy as $L/E_\nu$.  Since the relevant baselines are too short for oscillations from $\Delta m^2_{21}$ and $\Delta m^2_{31}$ mixing, the NC disappearance probability can be written as
\begin{equation}
  \begin{split}
    P(\nu_e\rightarrow\nu_s)=\sin^22\theta_{14}\cos^2\theta_{24}\cos^2\theta_{34}\sin^2\frac{\Delta m^2_{41}L}{4E_\nu} \\
    P(\nu_\mu\rightarrow\nu_s)=\cos^4\theta_{14}\sin^22\theta_{24}\cos^2\theta_{34}\sin^2\frac{\Delta m^2_{41}L}{4E_\nu}
  \end{split}
\end{equation}
for $\nu_e$ and $\nu_\mu/\bar{\nu}_\mu$ flavors, respectively.  With the multiple flavors produced at the SNS that are separable in time, our detectors can directly measure $\theta_{14}$, $\theta_{24}$, and $\Delta m^2_{14}$ from characteristic dips in our observed energy and time spectra. Oscillation probabilities depend weakly on $\theta_{34}$. This angle is known to be small from unitarity~~\cite{Denton:2021mso}, and we assume this parameter is 0.  

The disappearance channel is inherently favorable to study.  The $\nu_\mu\rightarrow\nu_e$ channel depends on the parameter $\sin^2\theta_{\mu e}=\sin^2\theta_{14}\sin^22\theta_{24}$, which is fourth order in the small angles $\theta_{i4}$.  The disappearance channels are only quadratic in $\theta_{i4}$.  Thus, though the LSND/MiniBooNE anomaly is only a $0.3\%$ effect, the disappearance channels each predict a $\approx 10\%$ effect at oscillation maximum.  Such a deficit would be detectable with precision CEvNS experiments.

\begin{figure}
  \includegraphics[width=0.48\textwidth]{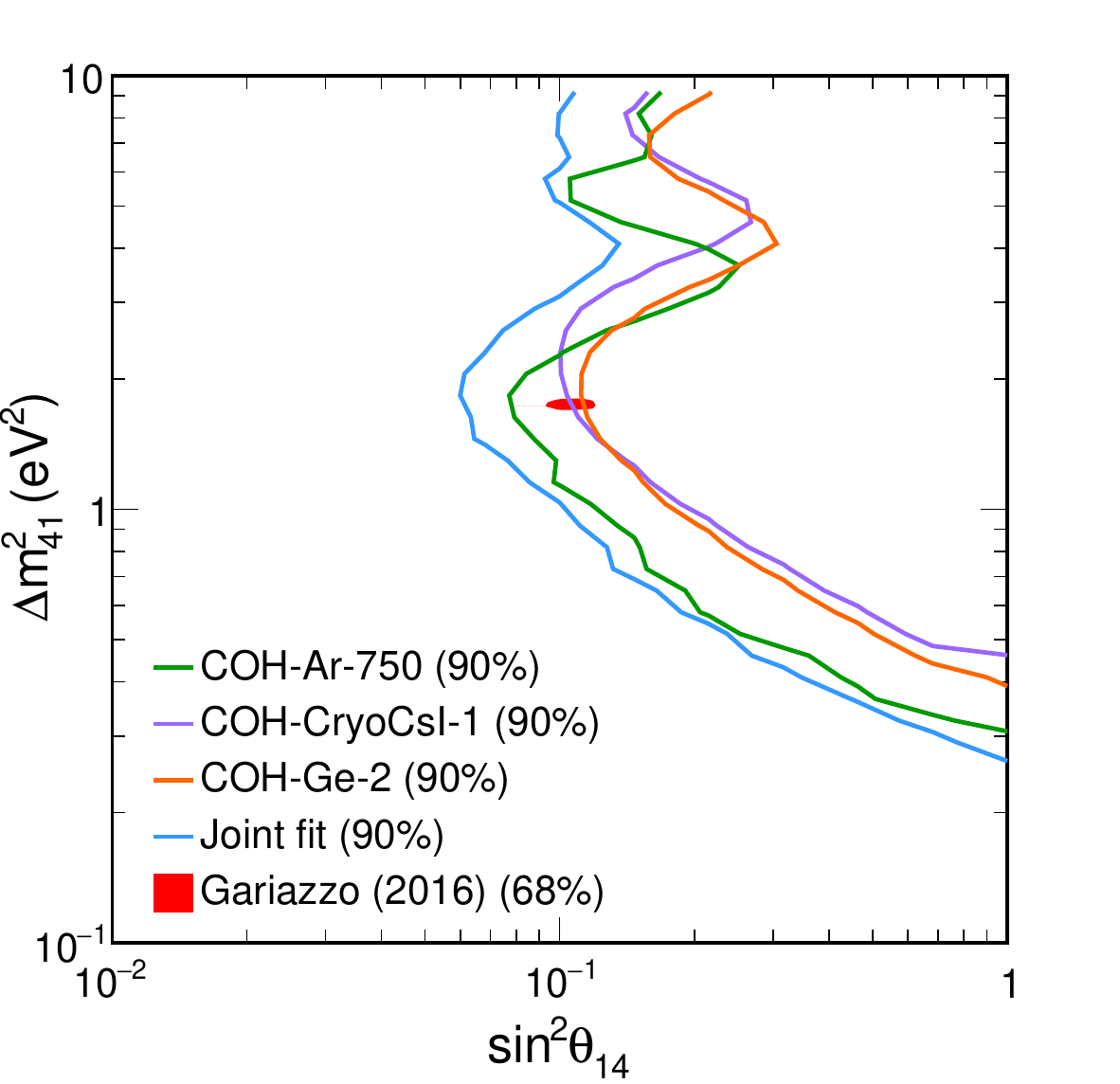}
  \includegraphics[width=0.48\textwidth]{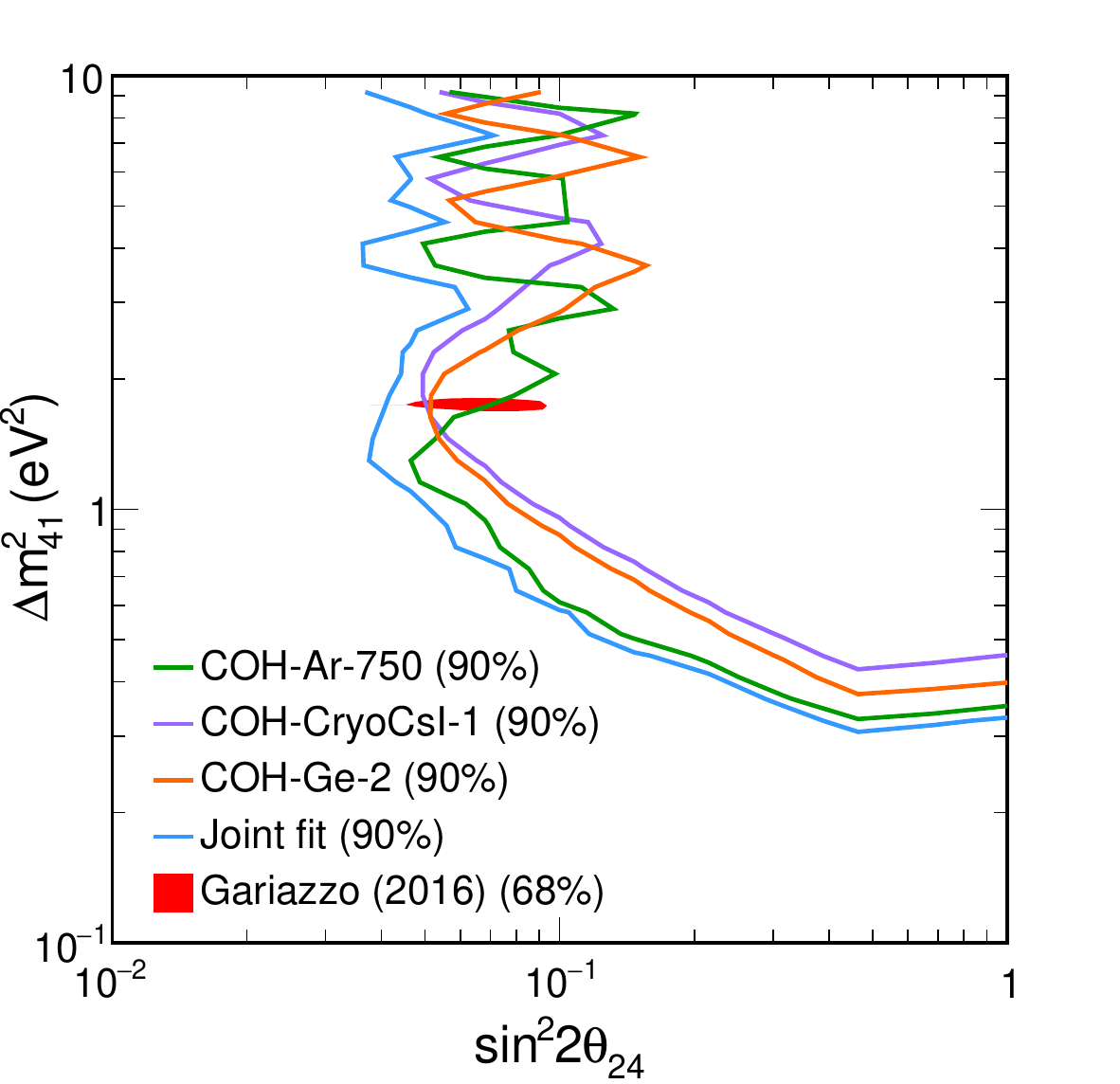}
  \caption{Sensitivities for COHERENT argon (detector to be completed 2024), cryogenic CsI (currently described), and germanium (upgrade with triple mass compared to detector currently running at the SNS) detectors to test the sterile-neutrino hypothesis in the $\nu_e$ (top) and $\nu_\mu$ (bottom) disappearance cases.  For each, a joint fit of all three datasets is compared against a global fit~\cite{Gariazzo:2017fdh} of all short-baseline oscillation data.}
  \label{fig:SterileSens}
\end{figure}

Beyond reduced interaction uncertainties with the CEvNS interaction channel, a $\pi$DAR neutrino flux gives a monoenergetic, prompt flux of $\nu_\mu$ at $E_\nu=29.8$~MeV.  Since the baseline is fixed, a measurement of the prompt CEvNS rate would precisely measure $\Delta m^2_{41}$ if a sterile state exists near the global fit.  CEvNS is a NC process, so there is only slight correlation between neutrino energy and observable recoil energy.  However, there is a maximum recoil energy of $2E_\nu^2/m_\mathrm{N}$, where $m_\mathrm{N}$ is the nuclear mass.  Thus, selecting the highest observable recoil energies also selects the highest energies produced at the SNS, effectively making a narrow-band near the flux endpoint.  Thus, with a 2D fit in recoil time and energy, COHERENT can test disappearance with $\nu_e$ and $\nu_\mu$ flavors and has two distinct signal regions with a narrow flux distribution.  Together with COHERENT's multiple-detector layout, a positive sterile-neutrino detection would have multiple cross-checks built into the analysis to distinguish between sterile oscillations and other new physics or systematic mismodeling.

In Fig.~\ref{fig:SterileSens}, we show expected sensitivities to $\Delta m^2_{41}$ and mixing angles after three years of running for three COHERENT detectors: COH-CryoCsI-1; COH-Ge-2, a future proposed concept of germanium PPC detector (50~kg); and COH-Ar-750.  For each detector, we assume three systematic uncertainties: 10$\%$ on the neutrino flux (which is correlated between all detectors), quenching (taken from CONUS~\cite{Bonhomme:2022lcz} for germanium and \cite{COHERENT:2020iec} for argon), and nuclear form factor (by varying $R_n\pm5\%$), Generally, the $\theta_{24}$ constraint is stronger as two flavors, $\nu_\mu/\bar{\nu}_\mu$ contribute.  The sensitivity depends sharply on $\Delta m^2_{41}$ above 4~eV$^2$ for the $\theta_{24}$ contour.  This is due to the monoenergetic $\nu_\mu$ flux which so sharply selects a specific $\Delta m^2_{41}$.  We also show a combined fit of all three COHERENT datasets which improves on each individual measurement due to improved understanding of the baseline dependence on the oscillation and cancellation of the correlated neutrino-flux uncertainty.  

The parameter space preferred by a 2016 global fit of sterile-neutrino data is also shown.  The fit prefers higher values of $\theta_{14}$ than $\theta_{24}$.  Given COHERENT's preferential sensitivity to $\theta_{24}$, COHERENT can probe the best-fit parameter space in both mixing angles.  For sterile neutrino oscillations, the combined information from multiple COHERENT sub-systems is very beneficial.  All detectors play an important role and cross-check each other by studying the $L/E_\nu$ oscillation dependence.  With the robust CEvNS signature, COHERENT is well positioned to shine new and valuable insight on this long-standing anomaly in the coming years.

\section{Measuring the neutron charge distribution}
\label{sect:Rn}

In the limit $Q^2=0$, the nucleus acts like a point source of weak charge, and CEvNS is truly coherent.  At finite momentum transfers where the deBroglie wavelength of the momentum transfer, $Q$, is not small compared to the nuclear radius, the scattering is only partially coherent.  In these situations, the CEvNS cross section is suppressed by the form factor, $\lvert F(Q^2)\rvert^2$ in Eqn.~\ref{eqn:CEvNSXSec}, which describes the spatial distribution of the weak nuclear charge.  This is currently the largest source of uncertainty on the standard-model prediction of the CEvNS cross section.  Conversely, the $Q^2$ dependence allows CEvNS experiments to directly measure the form factor~\cite{Amanik_2009,PhysRevC.86.024612,Payne:2019wvy,Reed:2020fdf} and access the nuclear equation of state.  As CEvNS is primarily sensitive to the neutron number of the nucleus, COH-CryoCsI-1 will primarily measure the neutron density distribution. To first order, this is determined by the neutron radius, defined as the root-mean-square distance to each neutron from the nucleus center, $R_n\equiv\sqrt{\langle R_n^2\rangle}$.


The nuclear physics of the weak charge distribution directly affects our understanding of the astrophysics of neutron stars~\cite{Piekarewicz:2022ycz}.  In heavy, neutron-rich nuclei like $^{133}$Cs and $^{127}$I, neutrons will extend beyond the proton distribution, forming a neutron skin given by the difference in neutron and proton radii, $R_n-R_p$.  The neutron skin relates to the surface tension which balances against the degeneracy pressure of the neutron matter~\cite{Horowitz:2000xj}.  The same nuclear physics determines the equation of state near the surface of a neutron star~\cite{Horowitz:2000xj,PhysRevLett.85.5296,Carriere_2003,Tsang:2012se}. As such, terrestrial scattering experiments directly clarify predictions of the radii of neutron stars.  This also includes the nuclear physics of binary neutron-star mergers~\cite{Baiotti:2019sew,Tsang:2019mlz,Fasano:2019zwm}, a very relevant area of work following the LIGO/Virgo detection of GW170817~\cite{PhysRevLett.119.161101}.  Terrestrial CEvNS experiments are also a vital component needed for also understanding the mass of neutron stars~\cite{Salinas:2023epr}.  In CsI specifically, measurements of $R_n$ improve measurements of the weak mixing angle at low $Q^2$ by reducing dominant uncertainties of atomic parity-violation measurements in $^{133}$Cs~\cite{Cadeddu:2018izq}.

\begin{figure}
  \includegraphics[width=0.48\textwidth]{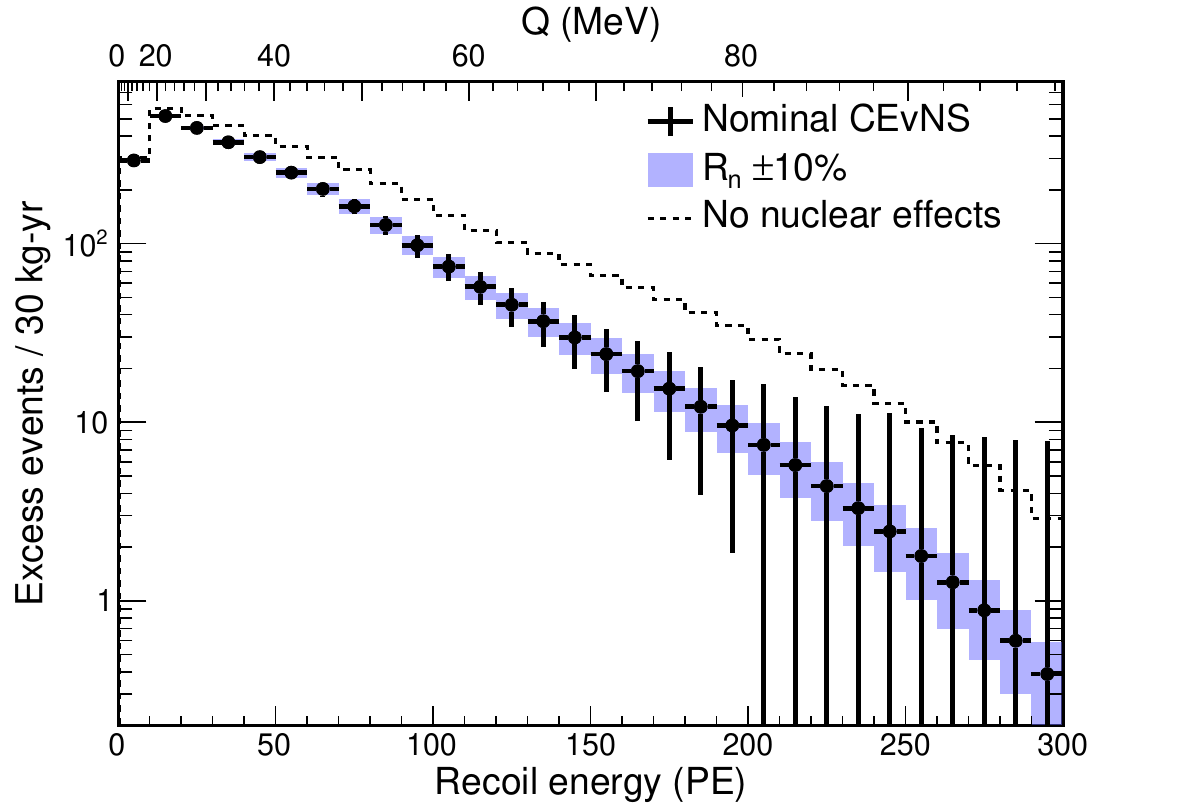}
  \caption{The expected signal CEvNS events, with backgrounds subtracted and an expected form-factor suppression from the Klein-Nystrand parameterization.  This is compared against the cases with $\pm10\%$ changes to the neutron radius and without form-factor suppression.}
  \label{fig:NuclearEffects}
\end{figure}


The weak form factor has been measured in the parity-violating electron scattering experiments, PREX~\cite{PhysRevLett.126.172502} and CREX~\cite{CREX:2022kgg}, which together saw a relatively large neutron skin in $^{208}$Pb and a small skin in $^{48}$Ca.  This is discrepant with nuclear models at $\approx2\sigma$~\cite{Reed:2023cap}.  Conveniently, precision CEvNS experiments are maturing in time to clarify these results.  Measurements with Ar in COH-Ar-750 will test the neutron skin with a light nucleus, like $^{48}$Ca, and CsI in COH-CryoCsI-1 will do the same in a heavier nucleus, analogous to $^{208}$Pb, and directly test whether the neutron skin increases with nuclear mass.

\begin{figure}
  \includegraphics[width=0.48\textwidth]{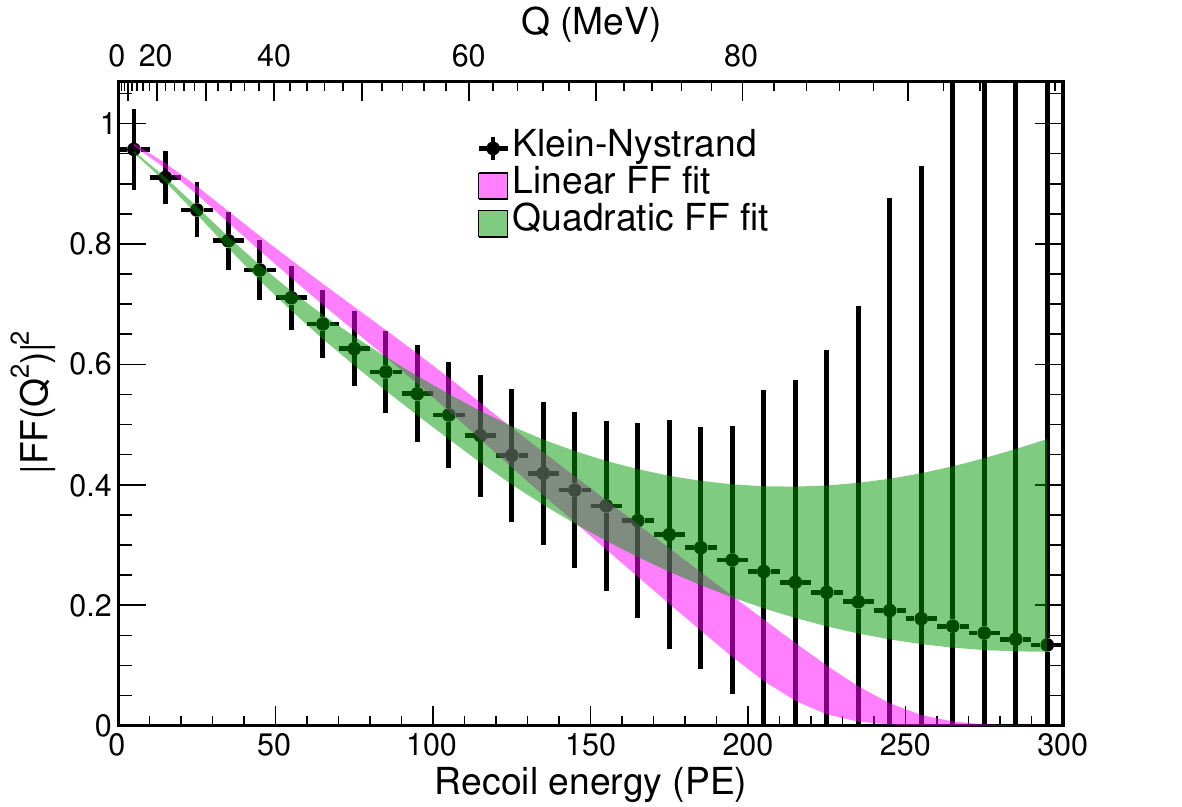}
  \caption{Reconstructed form factors, determined by the ratio of expected CEvNS scatters to the point-source expectation.  Additionally, the result is fit to a linear (purple) and quadratic (green) function.  The quadratic function, unlike the linear, can fully capture the $Q^2$ dependence at SNS energies.}
  \label{fig:PolyRn}
\end{figure}

\begin{figure}
  \includegraphics[width=0.48\textwidth]{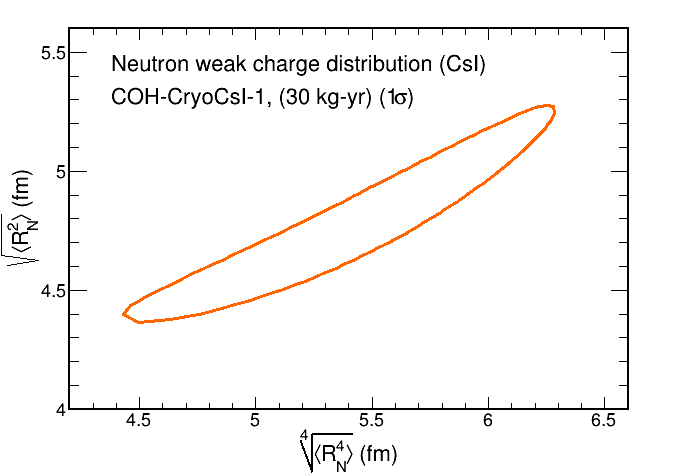}
  \caption{Sensitivity of COH-CryoCsI-1 to the linear and quadratic terms of the form-factor suppression, giving the neutron radius ($y-$axis) and $\sqrt[4]{\langle R_n^4\rangle}$ ($x-$axis).}
  \label{fig:PolyRnFit}
\end{figure}

To demonstrate the dependence of the CEvNS cross section on the nuclear equation of state, Fig.~\ref{fig:NuclearEffects} shows the CEvNS excess over background expected after 3~years of COH-CryoCsI-1 running compared to the CEvNS prediction with $R_n=0$.  The figure uses a larger uncertainty compared to nuclear physics calculations, $\pm10\%$ to illustrate the dependence on $R_n$.  Particularly at high recoil energies, and thus high $Q=\sqrt{2m_NE_\text{rec}}$, the nuclear effects are apparent.  To test the sensitivity of COH-CryoCsI-1 to $R_n$, we ran likelihood fits using fake data produced with the Klein-Nystrand form-factor parameterization~\cite{Klein:1999qj}.  For fitting, we did not assume any form-factor model, instead fitting the suppression to an arbitrary polynomial. Smearing between true recoil energy and observed PE was accounted for.  With the high light yield expected in the COH-CryoCsI-1 detector, smearing effects were minimal, smaller than the bin width for $Q<58$~MeV.  From~\cite{Amanik_2009,Sierra:2023pnf}, the linear term in the $Q^2$ Taylor expansion of the form factor directly relates to the neutron radius.  As shown in Fig.~\ref{fig:PolyRn}, next-generation CEvNS detectors like COH-CryoCsI-1 will also have sensitivity to the quadratic term in the expansion, which relates to $\sqrt[4]{\langle R_n^4\rangle}$, a measure of the diffuseness of the nuclear equation of state.  We can distinguish between quadratic and linear fits at a low statistical significance, $\approx 4\sigma$.


The determined sensitivity to both of these parameters is shown in Fig.~\ref{fig:PolyRnFit} after three years of running at the SNS.  Without assuming a form-factor parameterization, the detector could make a model-independent measurement of $R_n$ to $\approx7\%$ when profiling over $\sqrt[4]{\langle R_n\rangle}$.  There is $<1\sigma$ sensitivity to the cubic term in the $Q^2$ expansion with 10~kg of CsI, though this term may be accessible with an upgraded CsI detector.  Previous, a $2.9\%$ sensitivity to $R_n$ was calculated for COH-CryoCsI-1 that assumes a specific form-factor parameterization~\cite{Akimov:2022oyb}.  This level of precision is suitable for testing the large neutron skin observed in PREX~\cite{PhysRevLett.126.172502}, $\approx2\%$ uncertainty, in heavy nuclei.  


\section{Observing neutrinos from a galactic core-collapse supernova}

Detection of neutrinos from the supernova 1987a~\cite{1987IAUC.4316....1K,1989ARA&A..27..629A,Hirata:1988ad,HAINES198828,ALEXEYEV1988209} was pivotal for the development of neutrino astronomy and serves as an archetypical example of multi-messenger observation.  As a very massive star runs out of fusionable fuel, it undergoes a core-collapse supernova where its stellar core gravitationally collapses to either a neutron star or black hole.  In the process, the supernova releases $\sim10^{58}$ neutrinos with energies in the 10s of MeV over several seconds.  For supernovae within the Milky Way, this is a large enough flux to be detected in many neutrino experiments.  Due to the rarity of such collapses, $3\pm1$/century in the Milky Way, an observational approach utilizing multiple experiments with complementary sensitivity to the neutrino flux is desirable.  Several experiments currently running are actively waiting for the next galactic supernova neutrino burst~\cite{Super-Kamiokande:2016kji,NOvA:2020dll,IceCube:2014gqr,Bueno:2003ei,MicroBooNE:2020mqg,Lang:2016zhv,CADuba_2008} with more soon to turn on~\cite{DUNE:2020zfm,JUNO:2015zny,Hyper-Kamiokande:2018ofw,DarkSide20k:2020ymr}.  

As a NC process, CEvNS detectors are sensitive to all flavors of neutrinos and anti-neutrinos, giving a complete picture of the neutrino emission profile.  The NC channel is also insensitive to uncertainties in three-flavor neutrino oscillations, which become non-linear at neutrino densities experienced in the proto-neutron star~\cite{Duan:2005cp,Fogli:2007bk,Raffelt:2007cb}.  Given the large cross section for CEvNS, a reasonable event rate is possible with even ton-scale detectors.  Dark-matter experiments will observe CEvNS from a supernova with the large liquid-noble scintillation detectors driving sensitivity~\cite{Lang:2016zhv,DarkSide20k:2020ymr}.  Event rates expected in each detector are small, $\sim10-100$ CEvNS events for a collapse at 10~kpc, but given the multiple instrumented detectors throughout the world, the global CEvNS event rate can be significant.  

\begin{figure}
  \includegraphics[width=0.48\textwidth]{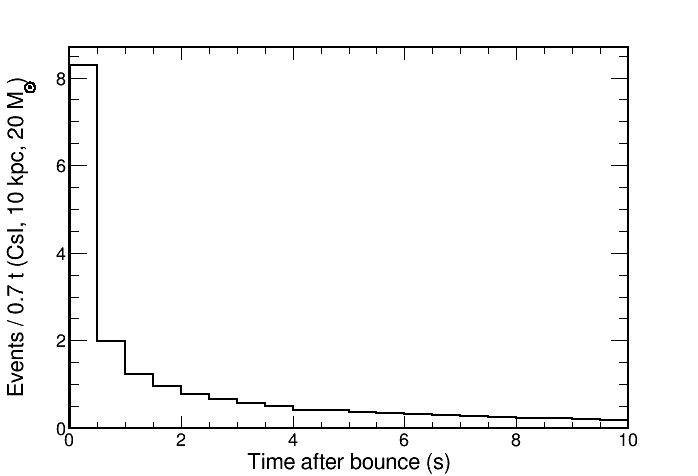}
  \caption{The CEvNS event rate per bin in 700~kg of cryogenic CsI for a representative supernova collapse 10~kpc from Earth.}
  \label{fig:supernova}
\end{figure}

A second-generation CryoCsI detector, either placed at the SNS or developed as a dedicated underground astroparticle experiment, would contribute to this measurement, and benefits from experience gained from running COH-CryoCsI-1 at the SNS.  We assume COH-CryoCsI-2 has a mass of 0.7~t which similarly achieves a light yield of 50~PE/keV$_\mathrm{ee}$.  We show here potential sensitivity to a supernova neutrino flux produced in a collapse simulation~\cite{Nakazato_2013}.  As a representative sample, we considered a 20$M_\odot$ collapse with metalicity $z=0.004$ and a shock revival time of 200~ms at a distance of 10~kpc.  This particular simulation provides information for 20~s following the supernova onset.  With the simulated efficiency estimated for 700~kg of instrumented CryoCsI, such a detector would expect 20.7 CEvNS interactions from this progenitor.  The time profile is shown in Fig.~\ref{fig:supernova}.  


At the ton-scale, CEvNS detectors can detect supernova neutrino bursts from across the galaxy with an event rate of $\sim1/$50~kg from a typical burst at 10~kpc.  COH-CryoCsI-1 alone would only be sensitive to supernovae within $\approx1$~kpc from Earth.  When combined with other COHERENT detectors, the additional active CsI mass would improve event yields expected from a further burst.  A self-trigger in event of a supernova would be developed for COH-CryoCsI-2 while remaining detectors would trigger on a supernova through SNEWS~\cite{articleSNEWS}.

\section{Conclusions}

An undoped, inorganic scintillation detector operated at cryogenic temperatures is an excellent candidate technology for studying low-energy nuclear recoil signals from CEvNS interactions.  The high light yield achieved in these crystals corresponds to low detector thresholds, $\approx$~$80$~eV$_\mathrm{ee}$ for CsI.  We studied the physics potential of a small, 10-kg undoped, cryogenic CsI detector at the SNS, called COH-CryoCsI-1.  Due to the improved threshold, this yields an order-of-magnitude improvement in event rate relative to the first COHERENT CsI detector without increasing the detector mass.  Further, as a heavy nuclear target, this detector would complement COHERENT's next CEvNS efforts which focus on the light sodium, argon, and germanium targets.  

This technology fundamentally expands the physics reach of CEvNS detectors, allowing NSI tests at lower mediator masses.  The COH-CryoCsI-1 detector would be an excellent probe of BSM physics.  It would resolve the LMA vs LMA-Dark question currently plaguing neutrino oscillations and test many dark-photon interpretations of $g-2$ results.  The high light yield also improves time resolution, relevant for searching for accelerator-produced dark matter, and energy resolution, favorable for testing the weak nuclear structure.  Distinct features in the recoil time and energy spectrum can distinguish between these BSM physics effects and, if no new physics is detected, measurements of weak nuclear structure would connect closely with current questions in theoretical nuclear physics.  At the ton scale, this technology would also be sensitive to CEvNS from a core-collapse supernova, observing $\approx1$~event~/~50~kg for a typical collapse at 10~kpc, making an impactful supernova measurement inclusive of all neutrino flavors.  In summary, COH-CryoCsI-1 would both resolve questions in nuclear physics and astrophysics and search for new physics in many well-motivated directions.


\section{Acknowledgements}

COHERENT collaborators thank Peter Denton and Julia Gehrlein for conversations on the interplay between NSIs and neutrino mixing.  We thank Adam Aurisano for thoughts about the sterile oscillation disappearance probability for NC interactions.  We also thank Jorge Piekarewicz for thoughts on the role of neutrino-scattering measurements in constraining nuclear structure.


The COHERENT collaboration acknowledges the generous resources provided by the ORNL Spallation Neutron Source, a DOE Office of Science User Facility, and thanks Fermilab for the continuing loan of the CENNS-10 detector. We also acknowledge support from the Alfred~P. Sloan Foundation, the Consortium for Nonproliferation Enabling Capabilities, the National Science Foundation, the Korea National Research Foundation (No. NRF 2022R1A3B1078756), and the U.S. Department of Energy, Office of Science. Laboratory Directed Research and Development funds from ORNL also supported this project. This work was performed under the auspices of the U.S. Department of Energy by Lawrence Livermore National Laboratory under Contract DE-AC52-07NA27344. This research used the Oak Ridge Leadership Computing Facility, which is a DOE Office of Science User Facility. The work was supported by the Ministry of Science and Higher Education of the Russian Federation, Project ``New Phenomena in Particle Physics and the Early Universe'' FSWU-2023-0073.

\bibliography{main.bbl}

\end{document}

%% file: authors-CryoCsI.tex
\newcommand{\Dukedesc}{\affiliation{Department of Physics, Duke University, Durham, NC, 27708, USA}}
\newcommand{\TUNLdesc}{\affiliation{Triangle Universities Nuclear Laboratory, Durham, NC, 27708, USA}}
\newcommand{\Mephidesc}{\affiliation{National Research Nuclear University MEPhI (Moscow Engineering Physics Institute), Moscow, 115409, Russian Federation}}
\newcommand{\ITEPnewadesc}{\affiliation{National Research Center  ``Kurchatov Institute'' , Moscow, 123182, Russian Federation }}
\newcommand{\UTKdesc}{\affiliation{Department of Physics and Astronomy, University of Tennessee, Knoxville, TN, 37996, USA}}
\newcommand{\USDdesc}{\affiliation{Department of Physics, University of South Dakota, Vermillion, SD, 57069, USA}}
\newcommand{\NCSUdesc}{\affiliation{Department of Physics, North Carolina State University, Raleigh, NC, 27695, USA}}
\newcommand{\Sandiadesc}{\affiliation{Sandia National Laboratories, Livermore, CA, 94550, USA}}
\newcommand{\Tuftsdesc}{\affiliation{Department of Physics and Astronomy, Tufts University, Medford, MA, 02155, USA}}
\newcommand{\ORNLdesc}{\affiliation{Oak Ridge National Laboratory, Oak Ridge, TN, 37831, USA}}
\newcommand{\UWdesc}{\affiliation{Center for Experimental Nuclear Physics and Astrophysics \& Department of Physics, University of Washington, Seattle, WA, 98195, USA}}
\newcommand{\LANLdesc}{\affiliation{Los Alamos National Laboratory, Los Alamos, NM, 87545, USA}}
\newcommand{\CNLdesc}{\affiliation{Canadian Nuclear Laboratories Ltd, Chalk River, Ontario, K0J 1J0, Canada}}
\newcommand{\IUdesc}{\affiliation{Department of Physics, Indiana University, Bloomington, IN, 47405, USA}}
\newcommand{\VTdesc}{\affiliation{Center for Neutrino Physics, Virginia Tech, Blacksburg, VA, 24061, USA}}
\newcommand{\NCCUdesc}{\affiliation{Department of Mathematics and Physics, North Carolina Central University, Durham, NC, 27707, USA}}
\newcommand{\NCSUnucengdesc}{\affiliation{Department of Nuclear Engineering, North Carolina State University, Raleigh, NC, 27695, USA}}
\newcommand{\CMUdesc}{\affiliation{Department of Physics, Carnegie Mellon University, Pittsburgh, PA, 15213, USA}}
\newcommand{\FSUdesc}{\affiliation{Department of Physics, Florida State University, Tallahassee, FL, 32306, USA}}
\newcommand{\WJCdesc}{\affiliation{Washington \& Jefferson College, Washington, PA, 15301, USA}}
\newcommand{\UFdesc}{\affiliation{Department of Physics, University of Florida, Gainesville, FL, 32611, USA}}
\newcommand{\Concorddesc}{\affiliation{Department of Physical and Environmental Sciences, Concord University, Athens, WV, 24712, USA}}
\newcommand{\SLACdesc}{\affiliation{SLAC National Accelerator Laboratory, Menlo Park, CA, 94025, USA}}
\newcommand{\Laurentiandesc}{\affiliation{Department of Physics, Laurentian University, Sudbury, Ontario, P3E 2C6, Canada}}
\newcommand{\SNUdesc}{\affiliation{Department of Physics and Astronomy, Seoul National University, Seoul, 08826, Korea}}
\author{P.S.~Barbeau}\Dukedesc\TUNLdesc
\author{V.~Belov}\Mephidesc\ITEPnewadesc
\author{I.~Bernardi}\UTKdesc
\author{C.~Bock}\USDdesc
\author{A.~Bolozdynya}\Mephidesc
\author{R.~Bouabid}\Dukedesc\TUNLdesc
\author{J.~Browning}\NCSUdesc
\author{B.~Cabrera-Palmer}\Sandiadesc
\author{E.~Conley}\Dukedesc
\author{V.~da Silva}\Tuftsdesc
\author{J.~Daughhetee}\ORNLdesc
\author{J.~Detwiler}\UWdesc
\author{K.~Ding}\USDdesc
\author{M.R.~Durand}\UWdesc
\author{Y.~Efremenko}\UTKdesc\ORNLdesc
\author{S.R.~Elliott}\LANLdesc
\author{A.~Erlandson}\CNLdesc
\author{L.~Fabris}\ORNLdesc
\author{M.~Febbraro}\ORNLdesc
\author{A.~Galindo-Uribarri}\ORNLdesc\UTKdesc
\author{M.P.~Green}\TUNLdesc\ORNLdesc\NCSUdesc
\author{J.~Hakenm\"uller}\Dukedesc
\author{M.R.~Heath}\ORNLdesc
\author{S.~Hedges}\altaffiliation{Also at: Lawrence Livermore National Laboratory, Livermore, CA, 94550, USA}\Dukedesc\TUNLdesc
\author{B.A.~Johnson}\IUdesc
\author{T.~Johnson}\Dukedesc\TUNLdesc
\author{A.~Khromov}\Mephidesc
\author{A.~Konovalov}\altaffiliation{Also at: Lebedev Physical Institute of the Russian Academy of Sciences, Moscow, 119991, Russian Federation}\Mephidesc
\author{E.~Kozlova}\Mephidesc
\author{A.~Kumpan}\Mephidesc
\author{O.~Kyzylova}\VTdesc
\author{J.M.~Link}\VTdesc
\author{J.~Liu}\USDdesc
\author{A.~Major}\Dukedesc
\author{K.~Mann}\NCSUdesc
\author{D.M.~Markoff}\NCCUdesc\TUNLdesc
\author{J.~Mattingly}\NCSUnucengdesc
\author{P.E.~Mueller}\ORNLdesc
\author{J.~Newby}\ORNLdesc
\author{N.~Ogoi}\NCCUdesc\TUNLdesc
\author{J.~O'Reilly}\Dukedesc
\author{D.S.~Parno}\CMUdesc
\author{D.~P\'erez-Loureiro}\CNLdesc
\author{S.I.~Penttila}\ORNLdesc
\author{D.~Pershey}\email{dpershey@fsu.edu}\FSUdesc
\author{C.G.~Prior}\Dukedesc\TUNLdesc
\author{J.~Queen}\Dukedesc
\author{R.~Rapp}\WJCdesc
\author{H.~Ray}\UFdesc
\author{O.~Razuvaeva}\Mephidesc\ITEPnewadesc
\author{D.~Reyna}\Sandiadesc
\author{G.C.~Rich}\TUNLdesc
\author{D.~Rudik}\altaffiliation{Also at: University of Naples Federico II, Naples, 80138, Italy}\Mephidesc
\author{J.~Runge}\Dukedesc\TUNLdesc
\author{D.J.~Salvat}\IUdesc
\author{J.~Sander}\USDdesc
\author{K.~Scholberg}\Dukedesc
\author{A.~Shakirov}\Mephidesc
\author{G.~Simakov}\Mephidesc\ITEPnewadesc
\author{W.M.~Snow}\IUdesc
\author{V.~Sosnovtsev}\Mephidesc
\author{M.~Stringer}\CNLdesc
\author{T.~Subedi}\Concorddesc
\author{B.~Suh}\IUdesc
\author{B.~Sur}\CNLdesc
\author{R.~Tayloe}\IUdesc
\author{K.~Tellez-Giron-Flores}\VTdesc
\author{Y.-T.~Tsai}\SLACdesc
\author{J.~Vanderwerp}\IUdesc
\author{E.E.~van Nieuwenhuizen}\Dukedesc\TUNLdesc
\author{R.L.~Varner}\ORNLdesc
\author{C.J.~Virtue}\Laurentiandesc
\author{G.~Visser}\IUdesc
\author{K.~Walkup}\VTdesc
\author{E.M.~Ward}\UTKdesc
\author{T.~Wongjirad}\Tuftsdesc
\author{Y.~Yang}\USDdesc
\author{J.~Yoo}\SNUdesc
\author{C.-H.~Yu}\ORNLdesc
\author{A.~Zaalishvili}\Dukedesc\TUNLdesc